\documentclass[letterpaper,12pt]{article}   
\usepackage{osajnl2} 
\usepackage[draft]{hyperref} 

\usepackage{cite}
\usepackage{amsmath,epsfig,multirow}
\usepackage{epstopdf,color,booktabs,comment}
\usepackage[caption=false,font=footnotesize]{subfig}
\usepackage{algorithm}
\usepackage{algorithmic}
\usepackage{amsmath,amsfonts,amsthm,amssymb}
\usepackage{cleveref}

\usepackage{steinmetz}

\usepackage{url}

\DeclareMathOperator*{\argmax}{arg\,max}
\DeclareMathOperator*{\argmin}{arg\,min}

\newcommand{\ak}[1]{\underline{a}^{+}_{1,(#1)}}

\newcommand{\aopt}[0]{\underline{a}_{\sf opt}}

\newcommand{\aoptsvdhat}[0]{\underline{a}_{\sf opt, svd}}
\newcommand{\aopteighat}[0]{\underline{a}_{\sf opt, eig}}
\newcommand{\aoptsdphat}[0]{\underline{a}_{\sf opt, sdp}}
\newcommand{\aoptsdphatr}[0]{\widetilde{\underline{a}}_{\sf opt, sdp}}
\newcommand{\aoptsvd}[0]{\underline{a}_{\sf svd}}

\newcommand{\tnorm}[0]{\tau_{\sf normal}}
\newcommand{\tequal}[0]{\tau_{\sf equal}}

\newcommand{\topt}[0]{\tau_{\sf opt}}

\newcommand{\ie}{i.e., }
\newcommand{\etal}{et al. }

\newcommand{\aleftp}[0]{\underline{a}^{+}_{1}}
\newcommand{\aleftm}[0]{\underline{a}^{-}_{1}}
\newcommand{\arightp}[0]{\underline{a}^{+}_{2}}
\newcommand{\arightm}[0]{\underline{a}^{-}_{2}}
\newcommand{\flipud}{{\sf flipud}}

\newcommand{\aequal}{\underline{a}_{\sf equal}}

\newcommand{\pv}[0]{\underline{\theta}}
\newcommand{\pvopt}[0]{\underline{\theta}_{\sf opt}}

\newcommand{\pmf}[0]{\underline{p}}

\newcommand{\tk}[1]{\underline{\theta}^{+}_{1,(#1)}}

\newcommand{\Trace}[0]{\textrm{Tr}}

\newcommand{\Imag}[0]{\mbox{Im}}
\newcommand{\Real}[0]{\mbox{Re}}

\begin{document}

\title{Backscatter analysis based algorithms for increasing transmission through highly-scattering random media using phase-only modulated wavefronts}


\author{Curtis Jin$^{*}$, Raj Rao Nadakuditi, Eric Michielssen and Stephen C. Rand}
\address{Dept. of EECS, University of Michigan, Ann Arbor, Michigan 48109-2122, USA}
\address{$^*$Corresponding author: jsirius@umich.edu}

\begin{abstract}
Recent theoretical and experimental advances have shed light on the existence of so-called  `perfectly transmitting' wavefronts with transmission coefficients close to $1$ in strongly backscattering random media. These perfectly transmitting eigen-wavefronts can be synthesized by spatial amplitude and phase modulation.

Here, we consider the problem of transmission enhancement using phase-only modulated wavefronts. Motivated by \textcolor{black}{biomedical} applications in which it is not possible to measure the transmitted fields, we develop physically realizable iterative and non-iterative algorithms for increasing the transmission through such random media using backscatter analysis.  {We theoretically show that, despite the phase-only modulation constraint, the non-iterative algorithms will achieve at least about $25 \pi \% \approx 78.5 \%$  transmission \textcolor{black}{with very high probability}, assuming that there is at least one perfectly transmitting eigen-wavefront and that the singular vectors of the transmission matrix obey a maximum entropy principle so that they are isotropically random. }

We numerically analyze the limits of phase-only modulated transmission in 2-D with fully spectrally accurate simulators and provide rigorous numerical evidence confirming  our theoretical prediction in random media with periodic boundary conditions that is composed of hundreds of thousands of non-absorbing scatterers. We show via numerical simulations that the iterative algorithms we have developed converge rapidly, yielding highly transmitting wavefronts using relatively few measurements of the backscatter field. Specifically, the best performing iterative algorithm yields $\approx 70\%$ transmission using just $15-20$ measurements in the regime where the non-iterative algorithms yield $\approx 78.5 \%$ transmission but require measuring the entire  modal reflection  matrix.  Our theoretical analysis and rigorous numerical results validate our prediction that phase-only modulation with a given number of spatial modes will yield higher transmission than amplitude and phase modulation with half as many modes.
\end{abstract}


\maketitle 

\section{Introduction}
Multiple scattering by randomly placed particles frustrates the  passage of light through `opaque' materials such as turbid water, white paint, and egg shells. Thanks to the theoretical work of Dorokhov \cite{dorokhov1982transmission}, \color{black} Barnes and Pendry \etal \cite{pendry1990maximal,barnes1991multiple}\color{black}, and others   \cite{mello1988macroscopic,beenakker2009applications}, as well as the breakthrough experiments of  Vellekoop and Mosk \cite{vellekoop2008phase,vellekoop2008universal} and others \cite{popoff2010measuring,kohlgraf2010transmission,shi2010measuring,kim2012maximal,van2011optimal,aulbach2011control,cui2011high,cui2011parallel,stockbridge2012focusing}, we now understand that even though a normally incident wavefront will barely propagate through a thick slab of such media \cite{ishimaru1999wave}, a small number of eigen-wavefronts exist that exhibit a transmission coefficient close to one and hence propagate through the slab  without significant loss.

In highly scattering random media composed of non-absorbing scatterers, these `perfectly transmitting' eigen-wavefronts are the right singular vectors of the modal transmission matrix with singular values (or transmission coefficients) close to $1$. Thus, if the modal transmission matrix were measured using the techniques described in  \cite{popoff2010measuring,kohlgraf2010transmission,shi2010measuring,kim2012maximal}, one could compute the pertinent singular vector and synthesize a highly transmitting eigen-wavefront via spatial amplitude and phase modulation. The task of amplitude and phase modulating an optical wavefront is not, however, trivial. Calibration and alignment issues prevent \textcolor{black}{the use of} two independent spatial light modulators in series that separately modulate the  signal amplitude and phase. A viable option is to use the innovative method developed by van Putten et al. in \cite{van2008spatial} for full spatial phase and amplitude control using a twisted nematic LCD combined with a spatial filter.

In a recent paper \cite{jin2012iterative,cjin2013}, we assumed that amplitude and phase modulation was feasible, and developed iterative, physically-realizable algorithms for synthesizing highly transmitting eigen-wavefronts using just a few measurements of the backscatter field. We showed that the algorithms converge rapidly and achieve $95\%$ transmission using about $5-10$ measurements. Our focus on constructing highly transmitting eigen-wavefronts by using the information in the backscatter field was motivated by \textcolor{black}{biomedical applications}, where it is often impossible to measure transmitted fields. \color{black} Our work will be most helpful in settings where it is desirable to increase the amount of light transmitted through an intervening scattering medium such as in photodynamic therapy where a photosynthesizing agent on exposure to light produces a form of oxygen that can kill neighboring (cancerous) cells \cite{dougherty1998photodynamic}. Another promising application is in photoacoustic imaging \cite{wang2010photoacoustic} which exploits the photoacoustic effect whereby light is converted into heat by absorbing scatterers and the subsequent thermoelasctic expansion produces wideband ultrasonic emissions which can be detected by ultrasonic transducers to form images.  Recent breakthrough works have taken this a step further by exploiting the photoacoustic effect to focus light within the medium \cite{kong2011photoacoustic,chaigne2014light}. Since the strength of the photoacoustic effect is proportional to the scattered light intensity at the light absorber, one might reasonably expect that algorithms, such as ours, that can increase the amount of transmitted light through a medium can help improve the penetration depth of photoacoustic imaging (or photoacoustic imaging guided focusing) by inducing stronger photoacoustic signals from deeper in the medium.
\color{black}

\begin{figure}[B]
\centering
\includegraphics[trim = 100 365 65 150, clip = true, width = 0.95\textwidth]{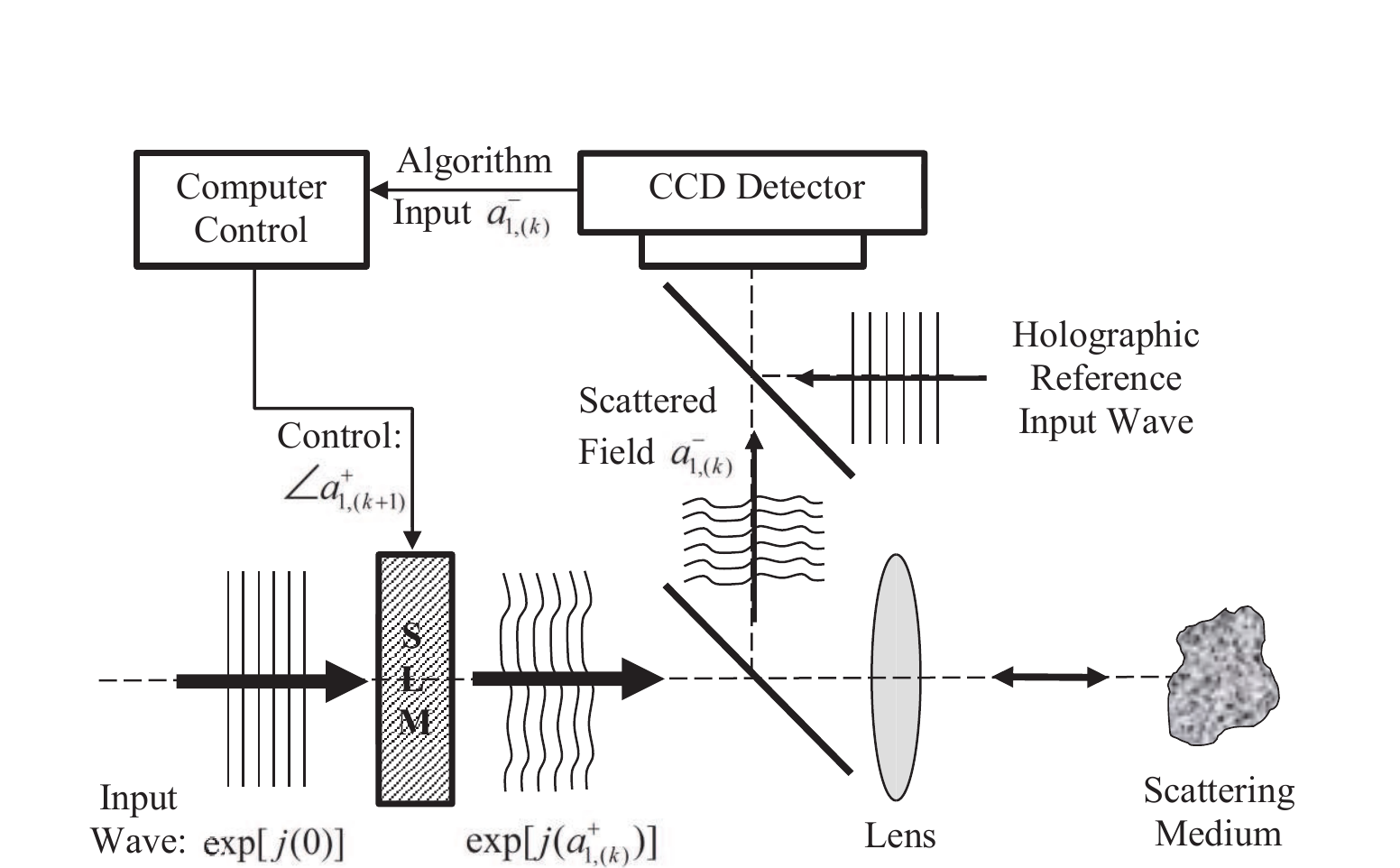}
\caption{Schematic for the experimental setup considered.}
\label{fig:exp setup}
\end{figure}

Here, we place ourselves in the setting where we seek to increase transmission via backscatter analysis but are restricted to phase-only modulation. The phase-only modulation constraint was initially motivated by the simplicity of the resulting experimental setup (see Fig. \ref{fig:exp setup}) and the commercial availability of finely calibrated phase-only spatial light modulators (SLMs) (e.g. the PLUTO series from Holoeye). As we shall shortly see there is another engineering advantage conferred by these methods. We do not, however, expect to achieve perfect transmission using phase-only modulation as is achievable by  amplitude and phase modulation. {However, we theoretically show that we can expect to get at least (about) $25 \pi \% \approx 78.5 \%$ provided that 1) the system modal reflection (or transmission) matrix is known, 2) its right singular vectors obey a maximum entropy principle by being isotropically random, and 3) full amplitude and phase modulation permits at least one perfectly transmitting wavefront.} We also develop iterative, physically realizable algorithms for transmission maximization that utilize backscatter analysis to produce a highly transmitting phase-only modulated wavefront in just a few iterations.  These rapidly converging algorithms build on the ideas developed in \cite{jin2012iterative,cjin2013} by incorporating the phase-only constraint. An additional advantage conferred by these rapidly converging algorithms is that they might facilitate their use in applications  where the duration in which the modal transmission or reflection matrix can be assumed to be constant is relatively small compared to the time it would take to make all measurements needed to estimate the modal transition or reflection matrix or in settings where a near-optimal solution obtained fast is  preferable to the optimal solution that takes many more measurements to compute. As in \cite{cjin2013}, the iterative algorithms we have developed retain the feature that they allow the number of modes being controlled via an SLM in experiments to be increased without increasing the number of measurements that have to be made.

We numerically analyze the limits of phase-only modulated transmission in 2-D with fully spectrally accurate simulators and provide rigorous numerical evidence confirming our theoretical prediction in random media with periodic boundary conditions that is composed of hundreds of thousands of non-absorbing scatterers. Specifically, we show that the best performing iterative algorithm yields $\approx 70\%$ transmission using just $15-20$ measurements in the regime where the non-iterative algorithms yield $\approx 78.5 \%$ transmission.

This theoretical prediction brings into sharp focus an engineering advantage to phase-only modulation relative to amplitude and phase modulation that we did not anticipate when we embarked on this line of inquiry. The clever idea in van Putten et al's work was to use spatial filtering to combine four neighboring pixels into one superpixel and then independently modulate the phase and the amplitude of light at each superpixel. This implies that an SLM with $M$ pixels can control at most $M/4$ spatial modes. For a given aperture, combining neighboring pixels into one super pixel corresponds to passing the entire wavefront into a spatial low pass filter. \textcolor{black}{We argue that in critically or undersampled scenarios phase-only measurements permit the design of more highly transmitting wavefronts than amplitude-phase measurements that use the idea of van Putten et al.}

 \color{black}  For highly scattering random media, our numerical results in  Section \ref{sec:simulations}, suggest that undersampling the spatial modes by $75\%$ will reduces the average amount of transmission by between $65 - 75 \%$. In contrast, our theoretical results in Section \ref{sec:funlimits} show that controlling all $M$ spatial modes using phase-only modulation will reduce the average amount transmission by at most $30\%$. Thus, we can, on average, achieve higher transmission with phase-only modulation using all the pixels in an SLM than by (integer-valued) undersampling of the pixels to implement amplitude and phase modulation!
\color{black}
The paper is organized as follows. We describe our setup in Section \ref{sec:Setup}. We discuss the problem of transmission maximization using phase-only modulated wavefronts in Section \ref{sec:Prob Form}.  We describe physically realizable, non-iterative and iterative algorithms for transmission maximization in Section \ref{sec:NonIterative} and in Section \ref{sec:AlgoBackMin}, respectively. We identify fundamental limits of phase-only modulated transmission in Section \ref{sec:funlimits}, validate the predictions and the rapid convergence behavior of the iterative algorithms in Section \ref{sec:simulations}, and summarize our findings in Section \ref{sec:conclusions}.

\section{Setup}
\label{sec:Setup}

\begin{figure}[B]
  \centering
  \includegraphics[trim = 0 18cm 13cm 9cm, clip=on]{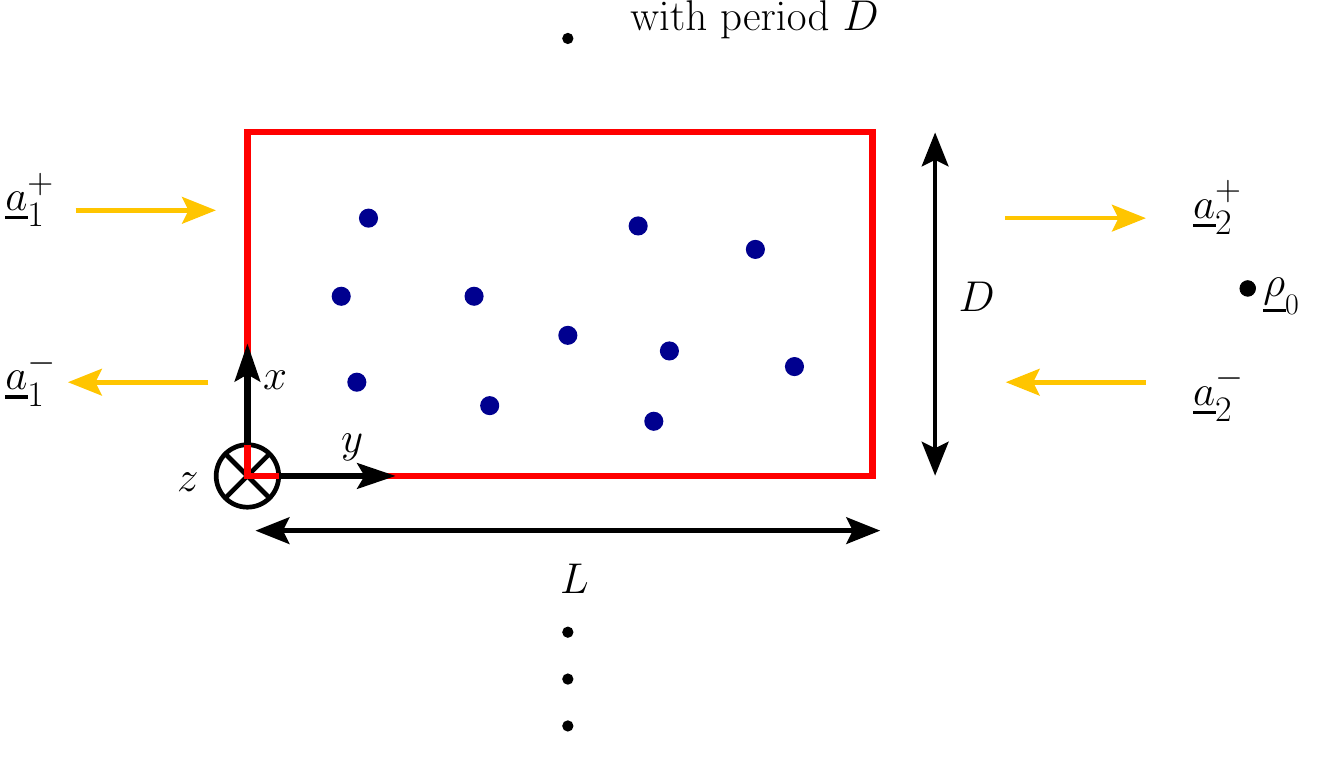}
  \caption{Geometry of the scattering system considered.}
  \label{fig:random model}
\end{figure}
We study scattering from a two-dimensional (2D) periodic slab of thickness $L$ and periodicity $D$. The slab's unit cell occupies the space $0 \leq x < D$ and $0 \leq y < L$ (Fig.  \ref{fig:random model}) and contains $N_{c}$ infinite and $z$-invariant circular cylinders of radius $r$ that are placed randomly within the cell and assumed either perfect electrically conducting (PEC) or dielectric with refractive index $n_{d}$. Care is taken to ensure the cylinders do not overlap. All fields are $\sf TM_{\mbox{$z$}}$ polarized: electric fields in the $y<0$ $(i=1)$ and $y>L$ $(i=2)$ halfspaces are denoted $\underline{e}_{i}(\underline{\rho})=e_{i}(\underline{\rho})\hat{z}$. These fields (complex) amplitudes $e_{i}(\underline{\rho})$ can be decomposed in terms of $+y$ and $-y$ propagating waves as $e_{i}(\underline{\rho}) = e_{i}^{+}(\underline{\rho}) + e_{i}^{-}(\underline{\rho})$, where
\begin{equation}
 \label{eq:IncidentWave}
e^{\pm}_{i}(\underline{\rho}) = \displaystyle \sum_{n=-N}^{N} h_{n} a^{\pm}_{i,n} e^{-j\underline{k}^{\pm}_{n} \cdot \underline{\rho}}\,.
\end{equation}
In the above expression, $\underline{\rho}=x\hat{x}+y\hat{y}\equiv(x,y)$, $\underline{k}^{\pm}_{n} = k_{n,x}\hat{x} \pm k_{n,y}\hat{y} \equiv (k_{n,x},\pm k_{n,y})$, $k_{n,x}=2\pi n/D$, $k_{n,y} = 2\pi \sqrt{(1/\lambda)^{2} - (n/D)^{2}}$, $\lambda$ is the wavelength, and $h_{n}=\sqrt{\| \underline{k}^{\pm}_{n} \|_{2} / k_{n,y}}$ is a power-normalizing coefficient. We assume $N=\lfloor D/\lambda \rfloor$, \ie we only model propagating waves and denote $M=2N+1$. The modal coefficients $a^{\pm}_{i,n}$, $i=1,2$; $n=-N,\ldots,N$ are related by the scattering matrix
\begin{equation}\label{eq:scat matrix}
\left[\begin{array}{c}\aleftm\\\arightp\\\end{array}\right] = \underbrace{\left[ \begin{array}{cc} S_{11} & S_{12} \\ S_{21} & S_{22} \end{array} \right]}_{=:S} \left[\begin{array}{c}\aleftp\\\arightm\\\end{array}\right],
\end{equation}
\color{black}where $\underline{a}^{\pm}_{i} = \begin{bmatrix} a^{\pm}_{i,-N} & \ldots & a^{\pm}_{i,-1} & a^{\pm}_{i,0} & a^{\pm}_{i,1} & \ldots & a^{\pm}_{i,N}\end{bmatrix}^{T}$ and $^{T}$ denotes transposition\color{black}. In what follows, we assume that the slab is only excited from the $y<0$ halfspace; hence, $\arightm=0$. For a given incident field amplitude $e^{+}_{1}(\underline{\rho})$, we define transmission and reflection coefficients as \begin{align}
\label{eq:tcoeff vec}
\tau(\aleftp) := \dfrac{\| S_{21}\cdot \aleftp \|_{2}^{2}}{\| \aleftp \|_{2}^{2}},\\
\intertext{and}
\label{eq:rcoeff vec}
\gamma(\aleftp) := \dfrac{\| S_{11}\cdot \aleftp \|_{2}^{2}}{\| \aleftp \|_{2}^{2}},
\end{align}
respectively. We denote the transmission coefficient of a normally incident wavefront by $\tnorm = \tau( \begin{bmatrix} 0 & \cdots &0 & 1 & 0 & \cdots &0 \end{bmatrix}^{T} )$. 

\section{Problem formulation}\label{sec:Prob Form}
 We define the phase-vector of the modal coefficient vector $\aleftp$, as
$$  \phase{\aleftp} = \begin{bmatrix} \quad \phase{a_{1,-N}^{+}} & \cdots & \phase{a_{1,0}^{+}} &  \cdots & \phase{a_{1,-N}^{+}} \quad  \end{bmatrix}^{T},$$
where for $n = -N, \ldots, N$,  $ a_{1,n}^+= |a_{1,n}^+| \exp(j  \phase{a_{1,n}^{+}})$ and $ |a_{1,n}^+|$ and $\phase{a_{1,n}^{+}}$ denote the magnitude and phase of $a_{1,n}^{+}$, respectively. \color{black} Let ${P}_{1}^{M}$ denote unit-norm vectors  of the form
\begin{equation}\label{eq:pvectors}
\underline{p}(\underline{\theta} ) = \sqrt{\dfrac{1}{M}} \begin{bmatrix} e^{j\theta_{-N}} & \cdots & e^{j\theta_{0}}  & \cdots & e^{j\theta_{N}} \end{bmatrix}^{T},
\end{equation}
where $\pv = \begin{bmatrix} \theta_{-N} & \cdots & \theta_{0}&  \cdots & \theta_{N} \end{bmatrix}^{T}$ is a $M \times 1$ vector of phases where $M:= 2N+1$. Then, the problem of designing a phase-only modulated incident wavefront that maximizes the transmitted power can be stated as
\begin{equation}\label{eq:optimization problem 1}
\underline{a}_{\sf opt} = \argmax_{ \aleftp \in P_{1}^M}  \tau(\aleftp) = \argmax_{ \aleftp \in P_{1}^{M} } \| S_{21} \cdot \aleftp \|_{2}^{2}.
\end{equation}
\color{black}

In the lossless setting, the scattering matrix $S$ in Eq. (\ref{eq:scat matrix}) will be unitary, \ie  $S^{H}\cdot S = I$, where $I$ is the identity matrix. Consequently, we have that
\begin{equation}\label{eq:unitary condition}
S_{11}^{H}\cdot S_{11} + S_{21}^{H}\cdot S_{21} = I,
\end{equation}
so that the $\| S_{21} \cdot \aleftp \|_{2} = \sqrt{1 - \| S_{11} \cdot \aleftp \|_{2}^{2}}$ and the optimization problem in Eq. (\ref{eq:optimization problem 1}) can be reformulated as
\begin{equation}\label{eq:bsmin}
\underline{a}_{\sf opt} = \argmin_{\aleftp \in P_{1}^M} \| S_{11} \cdot \aleftp \|_{2} =  \argmin_{\aleftp \in P_{1}^M}  \gamma(\aleftp).
\end{equation}
Thus the phase-only modulated wavefront that maximizes transmission will also minimize backscatter.
\color{black}
The feasible set in Eq. (\ref{eq:bsmin}) is non-convex since for $\theta_1 \neq \theta_2$ and $\alpha \in (0,1)$,  $\alpha\,\pmf(\underline{\theta}_1) + (1-\alpha) \pmf(\underline{\theta}_{2}) \notin P_{1}^{ M}$. Moreover, it is known \cite{smith1999optimum} that Eq. (\ref{eq:bsmin}) does not admit a closed-form solution for $\aopt$  (and hence $\pvopt$ ). Thus we turn our attention to computational methods for solving Eq. (\ref{eq:bsmin}).
\color{black}

\section{Non-iterative, phase-only modulating algorithms for transmission maximization}\label{sec:NonIterative}

We first consider algorithms for increasing transmission by backscatter minimization using phase-only modulated wavefronts that utilize measurements of the reflection matrix $S_{11}$. We assume that this matrix can be measured using the experimental techniques described in \cite{popoff2010measuring,kohlgraf2010transmission,shi2010measuring,kim2012maximal} by, in essence, transmitting $K > M$ incident wavefronts $\{ \underline{a}_{1,i}^+ \}_{i=1}^{K}$, measuring the (modal decomposition of the) backscattered wavefronts  $\{ \underline{a}_{1,i}^- \}_{i=1}^{K}$ and estimating $S_{11}$ by solving the system of equations  $\{ \underline{a}_{1,i}^- =  S_{11} \cdot  \underline{a}_{1,i}^+ \}_{i=1}^K$. We note that, even if the $S_{11}$ matrix has been measured perfectly, the optimization problem in Eq. (\ref{eq:bsmin})
 \textcolor{black}{is computationally intractable and known to be NP-hard \cite[Proposition 3.3]{zhang2006complex},\cite{luo2010semidefinite}.}
We can make the problem computationally tractable by relaxing the phase-only constraint in Eq. (\ref{eq:bsmin}) and allowing the elements of $\aleftp$ to take on arbitrary amplitudes and phases while imposing the power constraint $\parallel \aleftp \parallel^{2} = 1$. This yields the optimization problem
\begin{equation}\label{eq:prev optimization problem}
\aoptsvd=  \argmin_{\parallel \aleftp\parallel_{2} = 1} \| S_{11} \cdot \aleftp \|_{2}^{2},
\end{equation}
where we have relaxed the difficult constraint $\aleftp \in P_{1}^{M}$ into the spherical constraint $|| \aleftp||_{2} = 1$. The  problem in Eq. (\ref{eq:prev optimization problem}) can be solved exactly as described next.

Let $S_{21}= \sum_{i=1}^{M} {\sigma}_{i} \underline{{u}}_i \cdot \underline{{v}}_{i}^H$ and $S_{11}= \sum_{i=1}^{M} \widetilde{\sigma}_{i} \underline{\widetilde{u}}_i \cdot \underline{\widetilde{v}}_{i}^H$ denote the singular value decompositions (SVD) of $S_{21}$ and $S_{11}$, respectively. Here $\sigma_{i}$ (resp. $\widetilde{\sigma}_{i}$) is the singular value associated with the left and right singular vectors $\underline{u}_{i}$ and $\underline{v}_{i}$ (resp. $\underline{\widetilde{u}}_i$ and $\underline{\widetilde{v}}_i$), respectively. By convention, the singular values are arranged so that $\sigma_{1} \geq \ldots \geq \sigma_{M}$ and ${\widetilde{\sigma}}_{1} \geq \ldots \geq {\widetilde{\sigma}}_{M}$ and $^H$ denotes the complex conjugate transpose. \textcolor{black}{The solution to Eq. (\ref{eq:prev optimization problem})  can be expressed in terms of the right singular vectors of $S_{11}$ and $S_{21}$ as }
\begin{equation}\label{eq:aoptsvd}
\aoptsvd = \widetilde{\underline{v}}_{M} = \underline{v}_1.
\end{equation}
\textcolor{black}{In Eq. (\ref{eq:aoptsvd}), we have employed the well-known  variational characterization \cite[Theorem 7.3.10]{horn1990matrix}  of the smallest right singular vector for the first equality and the identity $\underline{v}_{i}  = \widetilde{\underline{v}}_{M-i+1}$ derived from Eq. (\ref{eq:unitary condition}) for  the second equality}.  This is an exact solution to the relaxed backscatter minimization problem in Eq. (\ref{eq:prev optimization problem}).

To get an approximation of the solution to the original unrelaxed problem in Eq. (\ref{eq:bsmin}) we construct a wavefront as
\begin{equation}\label{eq:aopt s21}
\aoptsvdhat = \pmf\left({\phase{\,\aoptsvd}}\right).
\end{equation}
The spherical relaxation that yields the optimization problem in Eq. (\ref{eq:prev optimization problem}) includes all the phase-only wavefronts in the original problem, but also includes many other wavefronts as well. We now consider a `tighter' semidefinite programming (SDP) relaxation that includes all the phase-only wavefronts in the original problem but fewer other wavefronts than the spherical relaxation does.

\color{black}

We note that SDP relaxations to computationally intractable problems such as Eq. (\ref{eq:bsmin}) have gained in popularity in recent decades because there are many problems in the literature for which the SDP relaxation is known to provide a constant relative accuracy estimate for the exact solution to the unrelaxed problem \cite{goemans1995improved,nesterov1998semidefinite,nesterov2000semidefinite}. We shall provide a similar constant relative accuracy estimate for our problem shortly in Eq. (\ref{eq:worstcase sdp}).

\color{black}

We begin by examining the objective function on the right hand side of Eq. (\ref{eq:prev optimization problem}). Note that
\begin{equation}\label{eq:sdp 1}
||S_{11} \cdot \aleftp||_{2}^2 = \left( (\aleftp)^{H} \cdot S_{11}^{H} \cdot S_{11} \cdot \aleftp \right) = \Trace \left( S_{11}^{H} \cdot S_{11} \cdot \aleftp \cdot (\aleftp)^{H} \right),
\end{equation}
where $\Trace(\cdot)$ denotes the trace of its matrix argument. Let us define a new matrix-valued variable $A = \aleftp \cdot (\aleftp)^H$. We note that $A$ is a Hermitian, positive semi-definite matrix with rank $1$ and $A_{ii} = 1/M$ whenever  $\aleftp \in P_{1}^{M}$, where $A_{ii}$ denotes the $i$th diagonal element of the matrix $A$.  Consequently, from  Eq. (\ref{eq:sdp 1}), we can derive the modified optimization problem
\begin{equation}\label{eq:sdp 2}
\begin{aligned}
{A}_{\sf opt} =& \argmin_{A \in \mathbb{C}^{M \times M}} \Trace\left( S_{11}^{H} \cdot S_{11} \cdot A \right)  \\
& \text{subject to } A = A^H, A \succeq 0, \textrm{rank}(A) = 1 \textrm{ and }  A_{ii} = 1/M \textrm{ for } i = 1, \ldots M, \\
\end{aligned}
\end{equation}
where the conditions $A = A^H$ and $A \succeq 0$ imply that $A$ is a Hermitian, positive semi-definite matrix. If we can solve Eq. (\ref{eq:sdp 2}) exactly, then by construction, since $A_{\sf opt}$ is rank $1$, we must have that $A_{\sf opt} = \aopteighat \cdot \aopteighat^H$ with $\aopteighat \in P_{1}^{M}$ so we would have solved Eq. (\ref{eq:bsmin}) exactly.  \textcolor{black}{Note that the set of rank one matrices is non-convex since the sum of two rank one matrices is not necessarily rank one. Thus the rank constraint in Eq. (\ref{eq:sdp 2}) makes the problem difficult to solve \cite{luo2010semidefinite} even though the objective function and other constraints are convex in $A$.}

Eliminating the difficult rank constraint yields the semi-definite programming (SDP) problem  \cite{vandenberghe1996semidefinite}
\begin{equation}\label{eq:SDP problem}
\begin{aligned}
A_{\sf sdp} =& \argmin_{A \in \mathbb{C}^{M \times M}} \Trace\left( S_{11}^{H} \cdot S_{11} \cdot A \right)  \\
& \text{subject to }   A= A^H, A \succeq 0, \textrm{ and }  A_{ii} = 1/M \textrm{ for } i = 1, \ldots M,\\
\end{aligned}
\end{equation}
which can be efficiently solved in polynomial-time \cite{luo2010semidefinite} using off-the shelf solvers such as CVX \cite{cvx, gb08} or SDPT3  \cite{SDPT3}. See Appendix \ref{sec:solvers} for details.

The computational cost of solving Eq. (\ref{eq:SDP problem}) and obtaining $A_{\sf sdp}$ is  $O(M^{4.5})$ \cite{luo2010semidefinite} while the computational cost for obtaining  $\aoptsvdhat$ using the Lanczos method for computing only the leading singular vector is $O(M^2)$ \cite{golub2012matrix}. Thus when $M > 1000$, there is a significant extra computational burden in obtaining the SDP solution. Hence, the question of when the extra computational burden of solving the SDP relaxation yields `large enough' gains relative to the spherical relaxation is of interest. We provide an answer using extensive numerical simulations in Section \ref{sec:simulations}.

We note that $A_{\sf sdp}$ is the solution to the relaxed backscatter minimization problem in Eq. (\ref{eq:SDP problem}). If $A_{\sf sdp}$ thus obtained has rank $1$ then we will have solved the original unrelaxed problem in Eq. (\ref{eq:bsmin}) exactly as well. Typically, however, the matrix $A_{\sf sdp}$ will not be rank one so we describe a procedure next for obtaining an approximation to the original  unrelaxed problem in Eq. (\ref{eq:bsmin}).

Let $A_{\sf sdp}= \sum_{i=1}^{M} \lambda_{i}\, \underline{u}_{i,{\sf sdp}} \cdot \underline{u}_{i,{\sf sdp}}^H$ denote the eigenvalue decomposition of $A_{\sf sdp}$ with the eigenvalues arranged so that $\lambda_{1} \geq \ldots \lambda_{M} \geq 0$. From $A_{\sf sdp}$, we can construct a phase modulated wavefront as
\begin{equation}\label{eq:aoptsdp}
\aoptsdphat = \pmf\left(\phase{\,\underline{u}_{1,{\sf sdp}}}\right).
\end{equation}
Since the SDP relaxation is a tighter relaxation than the spherical relaxation \cite{luo2010semidefinite}, we expect $\aoptsdphat$ to result in higher transmission than $\aoptsvdhat$. \textcolor{black}{Note that  $\aoptsdphat$ given by Eq. (\ref{eq:aoptsdp}) is an approximation to the solution of Eq. (\ref{eq:bsmin}). It is not guaranteed to be the phase-only modulated wavefront that yields the highest transmission. It does however provide a lower bound on the amount of transmission that can be achieved. }

\color{black}

In Eq. (\ref{eq:aoptsdp}) we constructed a deterministic approximation to $\aopt$ from $A_{\sf sdp}$. Consider the randomized approximation $\aoptsdphatr$ produced from $A_{\sf sdp}$ as
\begin{equation}\label{eq:aoptsdpr}
 \aoptsdphatr = \pmf\left(\phase{\left(\sum_{i=1}^{M} \sqrt{\lambda_{i}}\, \underline{u}_{i,{\sf sdp}} \cdot \underline{u}_{i,{\sf sdp}}^H\right)  \cdot z}\right),
\end{equation}
where $z = z_{R} + \sqrt{-1}\, z_{I}$ and $z_{R}$ and $z_{I}$ are $M \times 1$ i.i.d. random vectors that are normally distributed with mean zero and covariance $I_{M}/2$. From the results of Zhang and Huang \cite[Section 3.2]{zhang2006complex} and So et al. \cite[Corollary 1]{so2007approximating} it follows that in the lossless setting, due to the equivalence between Eq. (\ref{eq:optimization problem 1}) and Eq. (\ref{eq:bsmin}), we have that
\begin{equation}\label{eq:worstcase sdp}
\dfrac{\pi}{4} \tau(\aopt) \leq \mathbb{E}_{z}[\tau(\aoptsdphatr)] \leq \tau(\aopt) \leq 1.
\end{equation}

In other words, the wavefront $\aoptsdphatr$ is guaranteed to produce, on average, at least $78.54\%$ of the transmission that the optimal (unknown) wavefront $\aopt$ would produce. Eq. (\ref{eq:worstcase sdp}) quantifies the extent to which $\aoptsdphatr$ is suboptimal to $\aopt$. It provides no guarantee that the phase-only modulated wavefronts will be highly transmitting. We now provide a theoretical analysis of the transmitted power we can expect to achieve using these phase-only modulated wavefronts that will show that on average we can indeed expect them to be highly transmitting.
\color{black}

\section{Theoretical limit of phase-only modulated light transmission} \label{sec:funlimits}

 When the wavefront $\aoptsvd$ is excited, the optimal transmitted power is $ \topt :=\tau(\aoptsvd) = \sigma_{1}^{2}$.  Similarly, when  the wavefront associated with the $i$-th right singular vector $\underline{v}_{i}$ is transmitted, the transmitted power is $\tau(\underline{v}_{i}) = \sigma_{i}^{2}$, which we refer to as the transmission coefficient of the $i$-th eigen-wavefront of $S_{21}$.

The theoretical distribution \cite{dorokhov1982transmission,pendry1990maximal,barnes1991multiple,mello1988macroscopic,beenakker2009applications}
 of the transmission coefficients for lossless random media (referred to as the DMPK distribution)  has density given by \begin{equation}\label{eq:DMPK}
f(\tau) = \lim_{M \to \infty} \dfrac{1}{M} \sum_{i=1}^{M} \delta\left(\tau-\tau(\underline{v}_{i})\right) =  \dfrac{l}{2L} \dfrac{1}{\tau \sqrt{1-\tau}}, \qquad \textrm{ for } 4 \exp(-L/2l) \lessapprox \tau \leq 1.
\end{equation}
In Eq. (\ref{eq:DMPK}), $l$ is the mean-free path through the medium. This implies that in the regime where the DMPK distribution is valid, we expect $\tau(\aoptsvd) \approx 1$ so that (near) perfect transmission is possible using amplitude and phase modulation.  We now analyze the theoretical limit of phase-only modulation in the setting where the $S_{21}$ (or $S_{11}$) matrix has been measured and we have computed $\aoptsvdhat$ or $\aoptsdphat$  as in Eq. (\ref{eq:aopt s21}) and Eq. (\ref{eq:aoptsdp}), respectively. In what follows, we provide a lower bound on the transmission we expect to achieve in the regime where the DMPK distribution is valid.

We begin by considering the wavefront $\aoptsvdhat$ which yields a transmitted power given by
\begin{align}
\tau( \aoptsvdhat ) &= \tau(\pmf\left({\phase{\,\aoptsvd}}\right)  = \| S_{21} \cdot \pmf\left({\phase{\,\aoptsvd}}\right) \|_2^2  \\
 & = \| U \cdot \Sigma \cdot V^H \cdot \pmf\left({\phase{\,\aoptsvd}}\right) \|_{2}^{2}= \|\Sigma \cdot V^H \cdot \pmf\left({\phase{\,\aoptsvd}}\right) \|_{2}^{2},\label{eq:step2}
\end{align}
where we arrive at the last equality by exploiting the fact that $||U \cdot x ||_{2} = || x||_2$ for any unitary $U$.
Define $  \widetilde{\underline{p}}(\phase{\aoptsvd}) = V^H \cdot \pmf\left({\phase{\aoptsvd}}\right)$. Then from Eq. (\ref{eq:step2}), we have that
\begin{align}
\tau( \aoptsvdhat ) & = \|\Sigma \cdot \widetilde{\underline{p}}(\phase{\,\aoptsvd}) \|_{2}^{2} \\
 						        & = \sum_{i=1}^{M} \sigma^{2}_{i} \,  |\widetilde{p}_{i}(\phase{\,\aoptsvd})|^{2} \geq \sigma^2_{1} \, |\widetilde{p}_{1}(\phase{\,\aoptsvd})|^{2}.
\end{align}
In the DMPK regime, we have that $\sigma_{1}^2 \approx 1$ from which we can deduce that
\begin{equation}\label{eq:step3}
\tau( \aoptsvdhat )  \gtrsim |\widetilde{p}_{1}(\phase{\,\aoptsvd})|^{2}.
\end{equation}
From Eq. (\ref{eq:aoptsvd}), we have that $\aoptsvd = \underline{v}_{1} = \widetilde{\underline{v}}_{M}$ so that if
$$\underline{v}_{1}^H = \begin{bmatrix} |v_{1,1}|\,  e^{-j \phase{\, v_{1,1}}} & \ldots & |v_{1,M}| \,e^{-j \phase{\, v_{1,M}}} \end{bmatrix},$$
then
\begin{equation}
\widetilde{p}_{1}(\phase{\,\aoptsvd}) = \underline{v}_{1}^H \cdot \pmf(\,\phase{\,v_1}) = \dfrac{1}{\sqrt{M}} \sum_{i=1}^{M} |v_{1,i}|,
\end{equation}
and
\begin{equation}\label{eq:step4}
|\widetilde{p}_{1}(\phase{\,\aoptsvd})|^2 = \dfrac{1}{M} \sum_{i=1}^{M} |v_{1,i}|^2 + \dfrac{2}{{M}} \sum_{i < j} |v_{1,i}|\cdot |v_{1,j}|.
\end{equation}
\color{black}
Taking expectations on both sides of Eq. (\ref{eq:step4}) gives us
\begin{equation}\label{eq:step5}
\mathbb{E}[|\widetilde{p}_{1}(\phase{\,\aoptsvd})|^2] = \dfrac{1}{M} \sum_{i=1}^{M} \mathbb{E}[|v_{1,i}|^2] + \dfrac{2}{{M}} \sum_{i < j} \mathbb{E}[ |v_{1,i}|\cdot |v_{1,j}|].
\end{equation}

We now invoke the maximum-entropy principle as in \textcolor{black}{Pendry et al's} derivations \cite{pendry1990maximal,barnes1991multiple} and assume that the vector $\underline{v}_{1}$ is uniformly distributed on the unit hypersphere. Since the uniform distribution is symmetric, for any indices $i$ and $j$, we have that  $ \mathbb{E} \left[|v_{1,i}|^2  \right]= \mathbb{E} \left[|v_{1,1}|^2 \right]$ and $\mathbb{E}\left[|v_{1,i}| \cdot |v_{1,j}| \right]= \mathbb{E}\left[|v_{1,1}| \cdot |v_{1,2}|\right]$. Consequently Eq. (\ref{eq:step5}) simplifies to
\begin{equation}
\mathbb{E}[|\widetilde{p}_{1}(\phase{\,\aoptsvd})|^2 ] = \mathbb{E}\left[|v_{1,1}|^2 \right] + \frac{2 M (M-1)}{2 M} \, \mathbb{E}\left[ |v_{1,1}| \cdot |v_{1,2}| \right]
\label{eq:step7}
\end{equation}
Since $\| \underline{v}_{1} \|_{2}^2 = \sum_{i=1}^{M} |v_{1,i}|^2 = 1$, from symmetry considerations, we have that
\begin{equation}\label{eq:ev1sq}
\mathbb{E}\left[|v_{1,1}|^2 \right] = \dfrac{1}{M}.
\end{equation}
Substituting Eq. (\ref{eq:ev1sq}) into Eq. (\ref{eq:step7}) gives
\begin{equation}\label{eq:step8}
\mathbb{E}[|\widetilde{p}_{1}(\phase{\,\aoptsvd})|^2 ] = (M-1) \, \mathbb{E}\left[ |v_{1,1}| \cdot |v_{1,2}| \right] + \dfrac{1}{M}.
\end{equation}
A useful fact that will facilitate analytical progress  is that the distribution of the complex-valued random variables $v_{1,i}$ can be exactly characterized. Specifically, we have that \cite[Chap. 3a]{rao2009linear}
\begin{equation}\label{eq:dist equality}
v_{1,i} \overset{d}{=} \dfrac{g_{i}}{\sqrt{|g_{1}|^2 + \ldots + |g_{M}|^2}},
\end{equation}
where $\overset{{d}}{=} $ denotes equality in distribution and $g_i = x_{i} + \sqrt{-1} \, y_{i}$ and $x_{i}$ and $y_{i}$ are i.i.d. normally distributed variables with mean zero and variance $1$.  Let $r_{i} = |g_{i}|$. The random variables $|r_{i}|$ are i.i.d. Rayleigh distributed \cite{siddiqui1962some} with density given by
$$ f_{r_{i}}(r) = r\,e^{-\frac{r^2}{2}} \qquad \textrm{ for } r \geq 0.$$
The random variable $s_{3}  := \sqrt{ \sum_{i=3}^{M} r_{i}^{2}}$, by construction, is independent of $r_1$ and $r_2$ and is $\chi$ distributed with $2(M-2)$ degrees of freedom. It has density given by \cite[Section 11.3]{forbes2011statistical}
$$f_{s_{3}}(r) = \dfrac{2^{3-M} \cdot r^{2M -5} e^{-\frac{r^2}{2}}}{\Gamma(M-2)} \qquad \textrm{ for } r\geq 0.$$
The first term on the right hand side of Eq. (\ref{eq:step8}) can be expressed in terms of these intermediate variables as
\begin{align*}
\mathbb{E}\left[|v_{1,1}| \cdot |v_{1,2}| \right] &= \mathbb{E}\left[\dfrac{|g_{1}| \cdot |g_{2}|}{|g_1|^{2} + |g_{2}|^{2} + (|g_{3} |^{2} + \ldots + |g_{M}|^{2})}  \right]  = \mathbb{E}\left[\dfrac{r_1 \cdot r_2}{r_{1}^{2}+ r_{2}^{2} + s_{3}^2}  \right] \\
& = \int_{0}^{\infty} \int_{0}^{\infty}\int_{0}^{\infty}\dfrac{r_1 \cdot r_2}{r_{1}^{2}+ r_{2}^{2} + s_{3}^2}\,  r_1\,e^{-\frac{r_1^2}{2}}  \cdot r_2\,e^{-\frac{r_2^2}{2}} \cdot  \dfrac{2^{3-M} \cdot s_3^{2M - 5} e^{-\frac{s_3^2}{2}}}{\Gamma(M-2)}\, {\rm d}r_1 \,{\rm d}r_2 \, {\rm d} s_3  \\
& = \dfrac{2^{3-M}}{\Gamma(M-2)} \int_{0}^{\infty} \int_{0}^{\infty}\int_{0}^{\infty}\dfrac{r_1^2 \cdot r_2^2 \cdot s_3^{2M - 5} }{r_{1}^{2}+ r_{2}^{2} + s_{3}^2}\,  \,e^{-\frac{r_1^2+r_2^2+s_3^2}{2}}  \, {\rm d}r_1 \,{\rm d}r_2 \, {\rm d} s_3 .
\end{align*}
Let $r_1 = r \, \sin(t)\,\cos(p)$, $r_2 = r \, \sin(t)\, \sin(p)$ and $s_3 = r\,\cos(t)$. With these change of variables we have that
\begin{equation}\label{eq:piover4m}
\begin{split}
\mathbb{E}\left[|v_{1,1}| \cdot |v_{1,2}| \right] &=\dfrac{2^{3-M}}{\Gamma(M-2)}  \times \\
&  \int_{0}^{\frac{\pi}{2}} \int_{0}^{\frac{\pi}{2}}\int_{0}^{\infty}\dfrac{\cos^2(p) \,\left(r \cos(t)\right)^{2M} \sec(t) \, \sin^2(p) \, \tan^4(t)}{r^3}\,  \,e^{-\frac{r^2}{2}}  \, {\rm d}r \,{\rm d}t \, {\rm d} p  \\
& = \dfrac{\pi}{4M}.
\end{split}
\end{equation}
Substituting Eq. (\ref{eq:piover4m}) into Eq.(\ref{eq:step8}) gives us
\begin{equation}\label{eq:expected p1}
\mathbb{E}[|\widetilde{p}_{1}(\phase{\,\aoptsvd})|^2 ] = \dfrac{\pi}{4} + \dfrac{4-\pi}{4M}.
\end{equation}
Taking expectations on both sides  of Eq. (\ref{eq:step3})  and substituting Eq. (\ref{eq:expected p1}) into the right hand side yields the inequality
\begin{equation}\label{eq:lower bound M1}
\mathbb{E}[\tau(\aoptsvdhat)]  \gtrsim \dfrac{\pi}{4} +   \dfrac{4-\pi}{4M}.
\end{equation}
Since $\tau(\aoptsdphat) \geq \tau(\aoptsvdhat)$, Eq. (\ref{eq:lower bound M1}) yields the inequality
\begin{equation}\label{eq:lower bound M2}
\mathbb{E}[\tau(\aoptsdphat) ] \geq \mathbb{E}[\tau( \aoptsvdhat)]   \gtrsim \dfrac{\pi}{4} +   \dfrac{4-\pi}{4M}.
\end{equation}
Letting $M \to \infty$ on both sides on Eq. (\ref{eq:lower bound M2}) gives us
\begin{equation}\label{eq:aslb}
\lim_{M\to \infty}\mathbb{E}[\tau(\aoptsdphat)] \geq  \lim_{M\to \infty}\mathbb{E}[\tau(\aoptsvdhat)]  \gtrsim \dfrac{\pi}{4}.
\end{equation}

From Eq. (\ref{eq:aslb}) we expect to achieve at least $25\,\pi \%$ when the $S_{21}$ (or $S_{11}$) matrix has been measured and we compute the phase-only modulated wavefront using $\aoptsvdhat$ or $\aoptsdphat$. In contrast, amplitude and phase modulation yields (nearly) $100\%$ transmission; thus the phase-only modulation incurs an average loss of at most $~22\%$.

We now show that when $M$ is large and we are in the DMPK regime, with very high probability, we can expect to lose not much more than $22\%$ of the transmitted power relative to an amplitude and phase modulated wavefront. To that end we note that by the triangle inequality
$$ \sum_{i=1}^{M} | x_{i} + \delta_i| - \sum_{i=1}^{M} |x_{i}| \leq 1 \cdot \sum_{i=1}^{M} | \delta_i|.$$
This implies that $\sqrt{M}\, \widetilde{p}_{1}(\cdot)$ is a $1$-Lipschitz function of the argument. Under the assumption that $\aoptsvd = \underline{v}_{1}$ has uniform distribution on the unit hypersphere, from the results in \cite[Theorem 2.3 and Prop. 1.8]{ledoux2005concentration} it follows that there are positive constants $c$ and $C$ such that for $M$ large enough, and for all $\epsilon >0$
\begin{equation}\label{eq:concentration}
\mathbb{P}\left( \sqrt{M} |\widetilde{p}_{1}(\phase{\,\aoptsvd}) - \mathbb{E}[\widetilde{p}_{1}(\phase{\,\aoptsvd})] | \geq \epsilon\right)\leq C e^{\left(- c M \epsilon^{2}\right)}
\end{equation}
or equivalently, by setting $\epsilon \mapsto \sqrt{M} \, \epsilon$, that
\begin{equation}\label{eq:concentration 2}
\mathbb{P}\left( |\widetilde{p}_{1}(\phase{\,\aoptsvd}) - \mathbb{E}[\widetilde{p}_{1}(\phase{\,\aoptsvd})] | \geq \epsilon\right)\leq C e^{\left(- c M^2 \epsilon^{2}\right)}.
\end{equation}
Eq. (\ref{eq:concentration 2}) shows that we expect $|\widetilde{p}_{1}(\phase{\,\aoptsvd})$ and hence $|\widetilde{p}_{1}(\phase{\,\aoptsvd})|^2$ to be concentrated around its mean given by Eq. (\ref{eq:expected p1}). Thus, from Eq. (\ref{eq:step3}) we can conclude that as $M \to \infty$ we expect to transmit very close to $25\pi \%$ with very high probability.
\color{black}
\setlength{\heavyrulewidth}{0.1em}
\newcommand{\otoprule}{\midrule[\heavyrulewidth]}
\renewcommand*\arraystretch{1.5}

\section{Iterative, phase-only modulated algorithms for transmission maximization}
\label{sec:AlgoBackMin}
\color{black}
In Section \ref{sec:NonIterative} we described three non-iterative techniques for constructing approximations to $\aopt$ in Eq. (\ref{eq:bsmin}) via backscatter analysis that first require the $S_{11}$ to be measured and then compute $\aoptsvdhat$, $\aoptsdphat$ or $\aoptsdphatr$ using Eq. (\ref{eq:aopt s21}), Eq. (\ref{eq:aoptsdp}) and Eq. (\ref{eq:aoptsdpr}), respectively.

We now develop physically-realizable, iterative algorithms for increasing transmission by backscatter minimization that utilize significantly fewer measurements than the $O(M)$ measurements it would take to first estimate $S_{11}$ and subsequently construct $\aoptsvdhat$ or $\aoptsdphat$.  We note we do yet not have an theoretical guarantees that these iterative algorithms will indeed  converge rapidly and produce highly transmitting wavefronts. We provide evidence, in Section \ref{sec:simulations}, of their rapid convergence using results from numerical simulations.

\color{black}
\subsection{Steepest Descent Method}
We first consider an iterative method, based on the method of steepest descent, for finding the wavefront $\aleftp$ that minimizes the objective function $\|S_{11}\cdot \aleftp \|_{2}^{2}$. At this stage, we consider arbitrary vectors $\aleftp$ instead of phase-only modulated vectors $\aleftp \in P_{1}^{M}$.  The algorithm utilizes the negative gradient of the objective function to update the incident wavefront as
\begin{align} \label{SD:UpdateEq}
\underline{\tilde{a}}^{+}_{1,(k)} &= \ak{k} - \mu \left. \frac{\partial \| S_{11} \cdot \aleftp \|_{2}^{2} }{\partial \aleftp} \right|_{\aleftp = \ak{k}}  \\
&  = \ak{k} - 2 \mu S_{11}^{H}\cdot S_{11} \cdot \ak{k},  \label{SD:UpdateEq2}
\end{align}
where $\ak{k}$ represents the modal coefficient vector of the incident wavefront produced at the $k$-th iteration of the algorithm and $\mu$ is a positive stepsize.  If we renormalize $\underline{\tilde{a}}^{+}_{1,(k)}$ to have $||\underline{\tilde{a}}^{+}_{1,(k)}||_{2} = 1$, we obtain the iteration
\begin{equation}\label{eq:sd power}
 \ak{k+1} = \dfrac{(I - 2 \mu S_{11}^{H}\cdot S_{11}) \cdot \ak{k}}{ || (I - 2 \mu S_{11}^{H}\cdot S_{11}) \cdot \ak{k} ||_2}.
\end{equation}
\textcolor{black}{
Eq.  (\ref{eq:sd power}) is precisely the power iteration \cite[Algorithm 27.1]{trefethen1997numerical} on the matrix $(I - 2 \mu S_{11}^{H}\cdot S_{11})$. Thus \cite[Theorem 27.1]{trefethen1997numerical}, in the limit of $k \to \infty$, the incident wavefront $\ak{k+1}$ will converge to $\aoptsvd$, which is the largest eigenvector of $(I - 2 \mu S_{11}^{H}\cdot S_{11})$ provided $\widetilde{\sigma}_{1} > \widetilde{\sigma}_{2}$ and we select $\mu < 1/\left(\widetilde{\sigma}_{1}^{2}+\widetilde{\sigma}_{M}^{2}\right)$ . In the DMPK regime, $\widetilde{\sigma}^{2}_{M} = 1-\sigma^{2}_{1} \approx 0$ while  $\widetilde{\sigma}^{2}_{1} = 1-\sigma^{2}_{M} \approx 1$. Thus selecting $\mu \lesssim 1$ is justified. This iteration forms the basis for Algorithm \ref{alg:SD} which produces a highly transmitting wavefront by iterative refinement the wavefront $\ak{k+1}$.}

\begin{algorithm}[t]
\centering
\caption{Steepest descent algorithm for finding \textbf{$\aoptsvd$}}
\label{alg:SD}
\begin{algorithmic}[1]\label{alg}
\STATE Input: $\ak{0} = \mbox{ Initial random vector with unit norm}$
\STATE Input: $0 < \mu < 1/\left(\widetilde{\sigma}_{1}^{2}+\widetilde{\sigma}_{M}^{2}\right) = \mbox{step size}$
\STATE Input: $\epsilon = $ Termination condition
\STATE $k=0$
\WHILE{$\| S_{11} \cdot \ak{k} \|_{2}^{2} > \epsilon $}
\STATE $\underline{\tilde{a}}^{+}_{1,(k)} = \ak{k} - 2 \mu S_{11}^{H}\cdot S_{11} \cdot \ak{k}$
\STATE $\ak{k+1} = \underline{\tilde{a}}^{+}_{1,(k)} / \| \underline{\tilde{a}}^{+}_{1,(k)} \|_{2}$
\STATE $k = k + 1$
\ENDWHILE
\end{algorithmic}
\end{algorithm}

We now describe how the update equation given by Eq. (\ref{SD:UpdateEq2}) , which requires computation of the gradient $S_{11}^{H}\cdot S_{11} \cdot \ak{k}$, can be physically implemented even though we have not measured $S_{11}$ apriori.

Let $\flipud(\cdot)$ represent the operation of flipping a vector or a matrix argument upside down so that the first row becomes the last row and so on. Let $F = \flipud(I)$ where $I$ is the identity matrix, and let $^*$ denote complex conjugation. In our previous work \cite{cjin2013}, we showed that reciprocity of the scattering system implies that

\begin{equation}\label{eq:rec S11}
S_{11}^{H} = F\cdot S_{11}^{*} \cdot F,
\end{equation}
which can be exploited to make  the gradient vector $S_{11}^{H}\cdot S_{11} \cdot \ak{k}$ physically measurable. To that end, we note that Eq. (\ref{eq:rec S11}) implies that
\begin{equation}\label{eq:S11 breakup}
S_{11}^{H}\cdot \aleftm = F\cdot S_{11}^{*} \cdot F \cdot \aleftm = F \cdot (S_{11} \cdot (F\cdot (\aleftm)^{*}))^{*}.
\end{equation}
where $\aleftm = S_{11}\cdot \ak{k}$. Thus, we can physically measure $S_{11}^{H}\cdot S_{11} \cdot \ak{k}$, by performing the following sequence of operations and the accompanying measurements:
\begin{enumerate}
\item{Transmit $\ak{k}$ and measure the backscattered wavefront $\aleftm =  S_{11}\cdot \ak{k}$.}
\item{Transmit the wavefront obtained by time-reversing the wavefront whose modal coefficient vector is $\aleftm$ or equivalently transmitting the wavefront $F \cdot (\aleftm)^{*}$.}
\item{Measure the resulting backscattered wavefront corresponding to $S_{11} \cdot (F \cdot (\aleftm)^{*})$ and time-reverse it to yield the desired gradient vector $S_{11}^{H} \cdot S_{11} \cdot \ak{k}$ as shown in Eq. (\ref{eq:S11 breakup}).}
\end{enumerate}
The above represents a physically realizable scheme for measuring the gradient vector, which we proposed in \textcolor{black}{our previous paper \cite{cjin2013}}. Since time-reversal can be implemented using phase-conjugating mirror, we referred to this as the \textit{double phase-conjugating method}.

For the setting considered here, we have the additional physically-motivated restriction that all transmitted wavefronts $\aleftp \in P_{1}^{M}$. However, the wavefront $\aleftm$ can have arbitrary amplitudes and so will the wavefront obtained by time-reversing it (as in Step 2 above) thereby violating the phase-only modulating  restriction and making Algorithm \ref{alg:SD}, physically unrealizable. This is also why algorithms of the sort considered by others in array processing e.g. \cite{smith1999optimum} cannot be directly applied here.

This implies that even though Algorithm \ref{alg:SD} provably converges to $\aoptsvd$, it cannot be used to compute $\aoptsvdhat$ as in Eq. (\ref{eq:aopt s21}) because it is not physically implementable given the phase-only modulation constraint. To mitigate this problem, we propose modifying the update step in Eq. (\ref{SD:UpdateEq2}) to
\begin{equation} \label{SD:UpdateEq Phase}
\underline{\tilde{a}}^{+}_{1,(k)} = \pmf\left(\ak{k} - 2 \mu \overline{a} \,  S_{11}^{H}\cdot \pmf(\phase{S_{11} \cdot \ak{k}})\right),
\end{equation}
where $\overline{a}$ is chosen such that all magnitudes of modal coefficients of $\overline{a} \, \pmf(\phase{\aleftm})$ are set to the average magnitude of modal coefficients of $\aleftm$. Then, by applying Eq. (\ref{eq:rec S11}) as before, we can physically measure $ \overline{a} \,  S_{11}^{H}\cdot \pmf(\phase{S_{11} \cdot \ak{k}})$ by performing the following sequence of operations and the accompanying measurements:
\begin{enumerate}
\item{Transmit $\ak{k}$ and measure the backscattered wavefront $\aleftm =  S_{11}\cdot \ak{k}$.}
\item{Compute the scalar $\overline{a} = \displaystyle \dfrac{ \sum_{n=-N}^{N} | a_{1,n}^{-}| }{\sqrt{M}}$.}
\item{Transmit the (phase-only modulated) wavefront obtained by time-reversing the wavefront whose modal coefficient vector is $ \pmf(\phase{\aleftm})$.}
\item{Measure the resulting backscattered wavefront, time-reverse it, and scale it with $\overline{a}$ to yield the desired gradient vector.}
\end{enumerate}

This modified iteration in Eq. (\ref{SD:UpdateEq Phase})  leads to the algorithm in the left column of Table \ref{tab:SD} and its physical counterpart in the right column of Table \ref{tab:SD}. \textcolor{black}{We do not have a convergence theory for this algorithm; we propose selecting $\mu <1$ as before.}

\begin{table}[]
    \centering
    \begin{tabular}{ll}
   \otoprule
   \textbf{Vector Operation} & \textbf{Physical Operation}\\
   \hline
    $1: \quad \aleftm=S_{11}\cdot \ak{k}$ & $1: \quad  \ak{k} \xrightarrow{\mbox{ \tiny Backscatter \quad}} \underline{a}_{1}^{\tiny -}$\\
    $2: \quad \overline{a} = \displaystyle \dfrac{ \sum_{n=-N}^{N} | a_{1,n}^{-}| }{\sqrt{M}}$ & $2: \quad  \overline{a} = \displaystyle \dfrac{ \sum_{n=-N}^{N} | a_{1,n}^{-}| }{\sqrt{M}}$\\
    $3: \quad \aleftm \leftarrow \pmf(\phase{\aleftm})$ & $3: \quad \aleftm \leftarrow  \pmf(\phase{\aleftm})$\\
    $4: \quad \aleftp=F \cdot (\aleftm)^{*}$ & $4: \quad  \underline{a}_{1}^{\tiny -} \xrightarrow{\mbox{ \tiny PCM \quad}} \underline{a}_{1}^{\tiny +}$\\
    $5: \quad \aleftm=S_{11} \cdot \aleftp$ & $5: \quad  \aleftp \xrightarrow{\mbox{ \tiny Backscatter \quad}} \underline{a}_{1}^{\tiny -}$\\
    $6: \quad \aleftp=F \cdot (\aleftm)^{*}$ & $6: \quad  \underline{a}_{1}^{\tiny -} \xrightarrow{\mbox{ \tiny PCM \quad}} \underline{a}_{1}^{\tiny +}$\\
    $7: \quad \underline{\tilde{a}}^{+}_{1}=\ak{k}-2\mu \overline{a} \aleftp$ & $7: \quad  \underline{\tilde{a}}^{+}_{1}=\ak{k}-2\mu \overline{a} \aleftp$\\
    $8: \quad \ak{k+1}= \pmf( \phase{ \underline{\tilde{a}}^{+}_{1}} ) $ & $8: \quad \ak{k+1}= \pmf( \phase{ \underline{\tilde{a}}^{+}_{1}} ) $\\
  \otoprule
 \end{tabular}
 \caption{Steepest descent algorithm for refining a highly transmitting phase-only modulated wavefront. The first column represents vector operations. The second column represents the physical (or experimental) counterpart. The operation $\aleftm \longmapsto F\cdot (\aleftm)^{*}$ can be realized via the use of a phase-conjugating mirror (PCM). The algorithm terminates when \textcolor{black}{$|| \gamma(\ak{k+1}) - \gamma(\ak{k}) ||_{2} < \epsilon$, where $\epsilon$ is a preset threshold.}}
\vspace{-0.45cm}
 \label{tab:SD}
\end{table}

\subsection{Gradient Method}

The wavefront updating step for the algorithm described in Table \ref{tab:SD} first updates both the amplitude and phase of the incident wavefront (in Step 7) and then `projects it' onto the set of phase-only modulated wavefronts (in Step 8). We now develop a gradient-based method that only updates the phase of the incident wavefront. From Eq. (\ref{eq:bsmin}), the objective function of interest is $\|S_{11}\cdot \pmf(\pv) \|_{2}^{2}$ which depends on the phase-only modulated wavefront. The algorithm utilizes the negative gradient of the objective function with respect to the phase vector to update the phase vector of the incident wavefront as
\begin{equation}\label{GD:UpdateEq}
\tk{k+1} = \tk{k} - \sqrt{M} \mu \left. \frac{\partial \| S_{11} \cdot \pmf(\pv) \|_{2}^{2} }{\partial \pv} \right|_{\pv = \tk{k}},
\end{equation}
where $\tk{k}$ represents the phase vector of the wavefront produced at the $k$-th iteration of the algorithm and $\mu$ is a positive stepsize. In Appendix \ref{app:GDGradient}, we show that
\begin{equation} \label{GD:gradient}
 \frac{\partial \| S_{11} \cdot \pmf(\pv) \|_{2}^{2} }{\partial \pv} \bigg|_{\pv = \tk{k}} = 2 \Imag \left[ \mbox{diag}\{ \pmf(-\tk{k}) \} \cdot S_{11}^{H}\cdot S_{11}\cdot \pmf(\tk{k}) \right],
\end{equation}
where $\mbox{diag}\{ \pmf(-\tk{k}) \}$ denotes a diagonal matrix with entries $\pmf(-\tk{k})$ along its diagonal.
\color{black}
From Eq. (\ref{GD:gradient}), we have that
\begin{align*}
\left\| \frac{\partial \| S_{11} \cdot \pmf(\pv) \|_{2}^{2} }{\partial \pv} \bigg|_{\pv = \tk{k}} \right\|_{2} & = \left \| 2 \Imag \left[ \mbox{diag}\{ \pmf(-\tk{k}) \} \cdot S_{11}^{H}\cdot S_{11}\cdot \pmf(\tk{k}) \right] \right \|_{2}, \\
& \leq 2 \|\mbox{diag}\{ \pmf(-\tk{k}) \}\|_{2} \cdot \widetilde{\sigma}^{2}_{1} \leq 2 \dfrac{1}{\sqrt{M}} \cdot 1 = \dfrac{2}{\sqrt{M}}.
\end{align*}
This motivates our separation of the $\sqrt{M}$ factor from the stepsize in Eq. (\ref{GD:gradient}) since the resulting  $\mu$ can be chosen to be $O(1)$ and independent of $M$.
Substituting Eq. (\ref{GD:gradient}) into the right-hand side of Eq. (\ref{GD:UpdateEq}) yields the iteration
\begin{equation}\label{GD:UpdateEq2}
 \tk{k+1} = \tk{k} -2\sqrt{M}\mu \Imag \left[ \mbox{diag}\{ \pmf(-\tk{k}) \} \cdot S_{11}^{H}\cdot S_{11}\cdot \pmf(\tk{k}) \right].
\end{equation}
\color{black}
To evaluate the update Eq. (\ref{GD:UpdateEq2}), it is necessary to measure the gradient vector $S_{11}^{H}\cdot S_{11}\cdot \pmf(\tk{k})$. \color{black}For the same reason as in the steepest descent scheme, we cannot use double-phase conjugation introduced in our previous paper \cite{cjin2013} because of the phase-only modulating restriction. \color{black}Therefore, we propose modifying the update step in Eq. (\ref{GD:UpdateEq2}) to
\begin{equation} \label{GD:UpdateEq3}
\tk{k+1} = \tk{k} -2\sqrt{M}\mu \overline{a} \Imag \left[ \mbox{diag}\{ \pmf(-\tk{k}) \} \cdot S_{11}^{H}\cdot \pmf ( \phase{ S_{11}\cdot \pmf(\tk{k}) } ) \right],
\end{equation}
and we use the modified double-phase conjugation as
\begin{enumerate}
\item{Transmit  $\pmf(\tk{k})$ and measure the backscattered wavefront  $\aleftm = S_{11} \cdot \pmf(\tk{k})$;}
\item{Compute the scalar $\overline{a} = \displaystyle \dfrac{ \sum_{n=-N}^{N} | a_{1,n}^{-}| }{\sqrt{M}}$;}
\item{Transmit the phase-only modulated wavefront obtained by time-reversing the wavefront whose modal coefficient vector is $ \pmf(\phase{\aleftm})$;}
\item{Measure the resulting backscattered wavefront, time-reverse it, and scale it with $\overline{a}$ to yield the desired gradient vector.}
\end{enumerate}

The phase-updating iteration in Eq. (\ref{GD:UpdateEq3})  leads to the algorithm in the left column of Table \ref{tab:GA} and its physical counterpart in the right column of Table \ref{tab:GA}. \textcolor{black}{We do not have a convergence theory for this algorithm; we propose selecting the value of $\mu = O(1)$ which leads to fastest convergence by a line search.}

\begin{table}[]
    \centering
    \begin{tabular}{ll}
   \otoprule
   \textbf{Vector Operation} & \textbf{Physical Operation}\\
   \hline
    $1: \quad \aleftm=S_{11}\cdot \pmf(\tk{k})$ & $1: \quad  \pmf(\tk{k}) \xrightarrow{\mbox{ \tiny Backscatter \quad}} \underline{a}_{1}^{\tiny -}$\\
    $2: \quad \overline{a} = \displaystyle \dfrac{ \sum_{n=-N}^{N} | a_{1,n}^{-}| }{\sqrt{M}}$ & $2: \quad  \overline{a} = \displaystyle \dfrac{ \sum_{n=-N}^{N} | a_{1,n}^{-}| }{\sqrt{M}}$\\
    $3: \quad \aleftm \leftarrow \pmf(\phase{\aleftm} )$ & $3: \quad \aleftm \leftarrow \pmf(\phase{\aleftm} )$\\
    $4: \quad \aleftp=F \cdot (\aleftm)^{*}$ & $4: \quad  \underline{a}_{1}^{\tiny -} \xrightarrow{\mbox{ \tiny PCM \quad}} \underline{a}_{1}^{\tiny +}$\\
    $5: \quad \aleftm=S_{11} \cdot \aleftp$ & $5: \quad  \aleftp \xrightarrow{\mbox{ \tiny Backscatter \quad}} \underline{a}_{1}^{\tiny -}$\\
    $6: \quad \aleftp=F \cdot (\aleftm)^{*}$ & $6: \quad  \underline{a}_{1}^{\tiny -} \xrightarrow{\mbox{ \tiny PCM \quad}} \underline{a}_{1}^{\tiny +}$\\
    \hline
    \multicolumn{2}{c}{$7: \quad \tk{k+1} = \tk{k} -2\,\sqrt{M} \mu\, \overline{a}\, \Imag\left[ \mbox{diag}\{ \pmf(-\tk{k}) \} \cdot \aleftp \right]$}\\

  \otoprule
 \end{tabular}
 \caption{Gradient algorithm for transmission maximization. The first column contains the updating iteration in Eq. (\ref{GD:UpdateEq3}) split into a series of individual updates so that they may be mapped into their physical (or experimental) counterparts in the column to their right. The operation $\aleftm \longmapsto F\cdot (\aleftm)^{*}$ can be realized via the use of a phase-conjugating mirror (PCM). The algorithm terminates when \textcolor{black}{$|| \gamma\left(\pmf(\tk{k+1})\right) - \gamma\left((\pmf(\tk{k})\right) ||_{2} < \epsilon$, where $\epsilon$ is a preset threshold.}}
\vspace{-0.45cm}
 \label{tab:GA}
\end{table}

\section{Numerical simulations}
\label{sec:simulations}

To validate the proposed algorithms and the theoretical limits of phase-only wavefront optimization, we adopt the numerical simulation protocol described in \cite{cjin2013}. Specifically, we  compute the scattering matrices in Eq. (\ref{eq:scat matrix}) via a spectrally accurate, T-matrix inspired integral equation solver that characterizes fields scattered from each cylinder in terms of their traces expanded in series of azimuthal harmonics. As in \cite{cjin2013}, interactions between cylinders are modeled using 2D periodic Green's functions. The method constitutes a generalization of that in \cite{mcphedran1999calculation}, in that it does not force cylinders in a unit cell to reside on a line but allows them to be freely distributed throughout the cell.  As in \cite{cjin2013}, all periodic Green's functions/lattice sums are rapidly evaluated using a recursive Shank's transform using the methods described in \cite{singh1991use,sidi2003practical}.  Our method exhibits exponential convergence in the number of azimuthal harmonics used in the description of the field scattered by each cylinder.  As in \cite{cjin2013}, in the numerical experiments below, care was taken to ensure 11-digit accuracy in the entries of the computed scattering matrices.

\color{black}
We now describe how the simulations were performed. We generated a random scattering system with $D=197\lambda, r=0.11\lambda, \widetilde{L} = 3.4 \times 10^5 \lambda, N_{c} = 430,000, n_{d} = 1.3$ and $M = 395$. The locations of the scatterers were selected randomly and produced a system with $\overline{l}=6.69\lambda$, where $\overline{l}$ is the average distance to the nearest scatterer. Let $L$ denote the thickness of the scattering system we are interested in analyzing. We vary $L$ from $\lambda$ to $\widetilde{L}$ and for each value of $L$ we compute the scattering matrices associated with only the scatterers contained in the $(0,L)$ portion of the $(0,\widetilde{L})$ system we have generated. This construction ensures that the average density per ``layer'' of the medium is about the same. We computed the reported statistics by simulating $1700$ random realizations of the scattering system.

\begin{figure}
\centering
\includegraphics[trim = 60 5 60 10, clip = true,width=\textwidth]{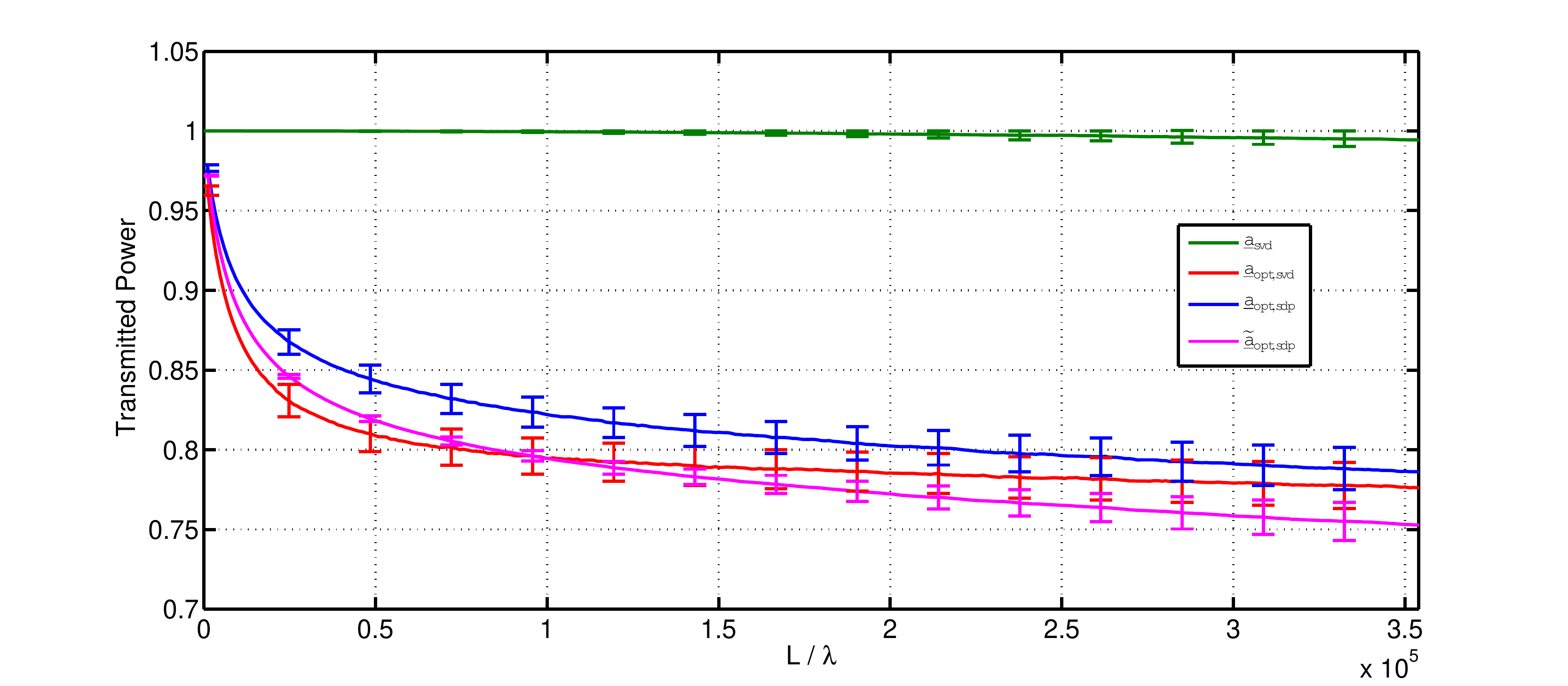}
\caption{Plot of transmitted power obtained by the SVD and SDP based algorithms versus  $L/\lambda$ in a setting where $D=197\lambda, r=0.11\lambda, n_{d} = 1.3$ and $M = 395$. The system was generated so that when $L = 3.4 \times 10^5 \lambda$, $N_{c} = 430,000$ and  $\overline{l}=6.69\lambda$, where $\overline{l}$ is the average distance to the nearest scatterer. The empirical average and the one-standard-deviation error bars were computed  over $1700$ random realizations of the scattering medium.}\vspace{-0.05cm}
\label{fig:SVDSDP}
\end{figure}

First we compare the transmitted power achieved by the non-iterative algorithms that utilize knowledge of the entire $S_{11}$ matrix to compute the wavefronts $\aoptsvdhat$,  $\aoptsdphat$ and $\aoptsdphatr$ given by Eq. (\ref{eq:aopt s21}),  Eq. (\ref{eq:aoptsdp}) and Eq. (\ref{eq:aoptsdpr}), respectively. Fig. \ref{fig:SVDSDP} compares the transmitted power for the SVD and SDP based algorithms as a function of the thickness  $L/\lambda$ of the scattering system averaged over $1700$ random realizations of the scattering system.

\color{black}
As expected, the wavefront $\aoptsdphat$ realizes increased transmission relative to the wavefront $\aoptsvdhat$. However, as the thickness of the medium increases, the gain vanishes. Typically $\aoptsdphat$ increases transmission by about $1-5\%$ relative to $\aoptsvdhat$. The wavefront $\aoptsdphatr$ is clearly suboptimal. Fig. \ref{fig:SVDSDP} also shows the accuracy of our theoretical prediction of $25 \, \pi \,\%  \approx 78.5 \%$ transmission using phase-only modulation for highly backscattering (or thick) random media in the same regime where the DMPK theory predicts perfect transmission using amplitude and phase modulated wavefronts. \color{black} The relatively small one-standard-deviation error bars displayed validate the prediction based on Eq. (\ref{eq:concentration 2}).

Recall that the computational cost of computing $\aoptsdphat$ is $O(M^{4.5})$  while the cost for computing $\aoptsvdhat$ is  $O(M^2)$. Fig.  \ref{fig:SVDSDP} suggests that for large $M$, the significantly extra computational effort  for computing $\aoptsdphat$ might not be worth the effort for strongly scattering random media.

\begin{figure}
\centering
\includegraphics[trim = 60 5 60 10, clip = true,width=\textwidth]{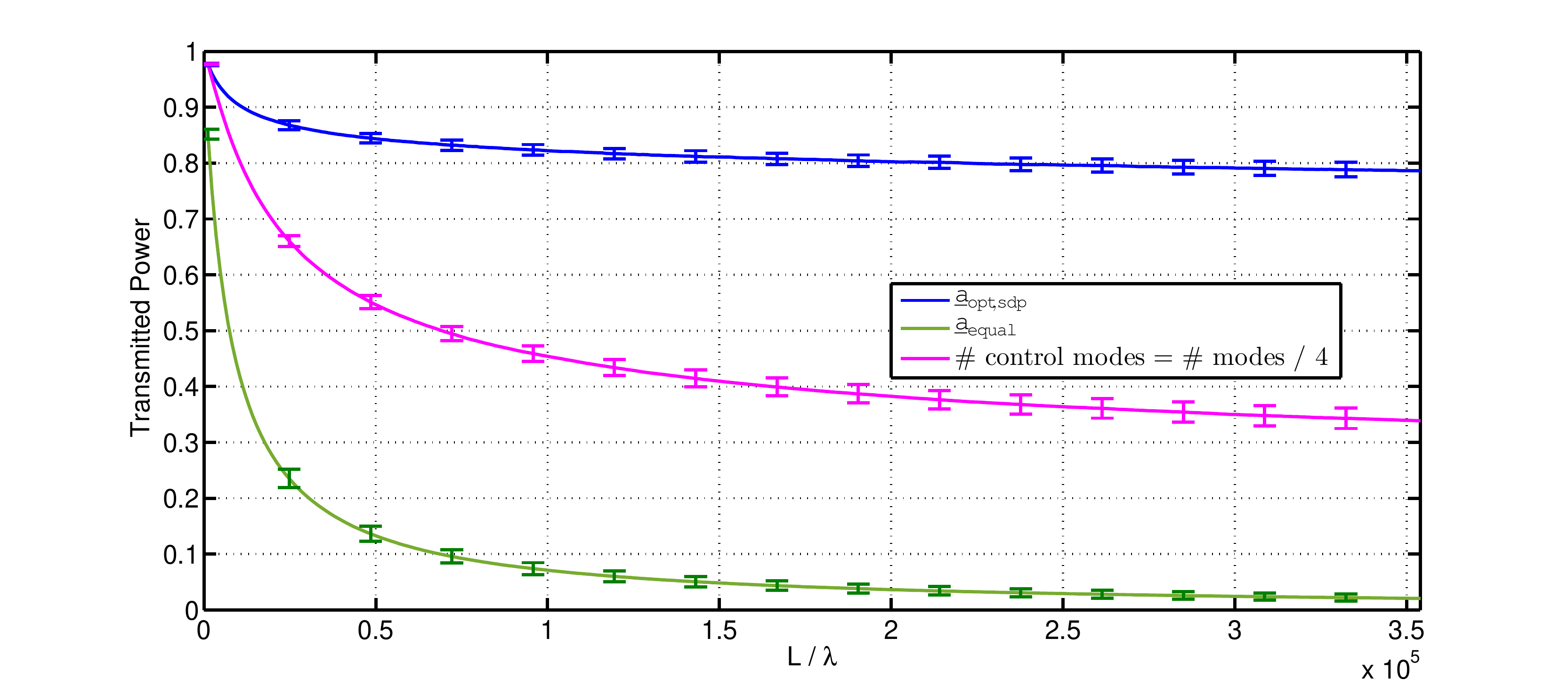}
\caption{Plot of transmitted power obtained by the $\aoptsdphat$, $\aequal$ and the amplitude and phase modulated wavefront corresponding to the largest right singular vector of the undersampled (by four) modal transmission transmission matrix versus $L/\lambda$ for the same setup as in Fig. \ref{fig:SVDSDP}.}\vspace{-0.05cm}
\label{fig:SVDSDP2}
\end{figure}

In Fig. \ref{fig:SVDSDP2}, we plot the transmitted power achieved by undersampling the number of control modes by a factor of $4$, computing the resulting $S_{21}$ matrix, and constructing the amplitude and phase modulated eigen-wavefront associated with the largest right singular vector. This is what would happen if we were to implement the `superpixel'-based amplitude and phase modulation scheme described in  \cite{van2008spatial} in the framework of a system with periodic boundary conditions. As can be seen, phase-only modulation yields higher transmission than amplitude and phase modulation with undersampled modes. We are presently studying whether the same result holds true in systems without periodic boundary conditions as considered in \cite{choi2011transmission}. 

Let $\aequal =  \underline{p}( \begin{bmatrix} 0 & \cdots & 0 & \cdots & 0 \end{bmatrix})$ represent a wavefront with equal phases (set arbitrarily to zero).  Fig.  \ref{fig:SVDSDP2} also plots the transmitted power achieved by the wavefront $\aequal$. The plot reveals that both the SVD and the SDP based algorithms realize significant gains relative to this vector\footnote{A normally incident wavefront also yields about the same transmitted power. Note that a normally incident wavefront cannot be synthesized using phase-only modulation using the setup in Fig. \ref{fig:exp setup}.}.

{\color{black} We shall now illustrate the performance of the iterative methods. For the iterative methods, let us denote the wavefront vector produced by the algorithm at the $k$-th iteration with stepsize $\mu$ as $\underline{a}_{1,(k)}^{+,\mu}$. In the simulations that follow, we chose the optimal $\mu$ for each algorithm, for every realization of the scattering medium, by computing\begin{equation}
  \mu_{\sf opt} = \argmax_{\mu_{\sf min} \leq \mu \leq \mu_{\sf max}} \sum_{k=0}^{50} \tau\left( \underline{a}_{1,(k)}^{+,\mu} \right).
\end{equation}}
In other words, the optimal $\mu$ was obtained by a line search, {i.e.}, by running the algorithms over a fixed set of discretized values of $\mu$ between $\mu_{\sf min}$ and $\mu_{\sf max}$, and choosing the $\mu$ that converged the fastest. In our experiments, we set $\mu_{\sf min} = 0.001$ (resp. $0.001$) and $\mu_{\sf max} = 1$ (resp. $5$)  for the steepest descent (resp. gradient descent) algorithm.

\begin{figure}
\centering
\includegraphics[trim = 0 10 0 25, clip = true,width=\textwidth]{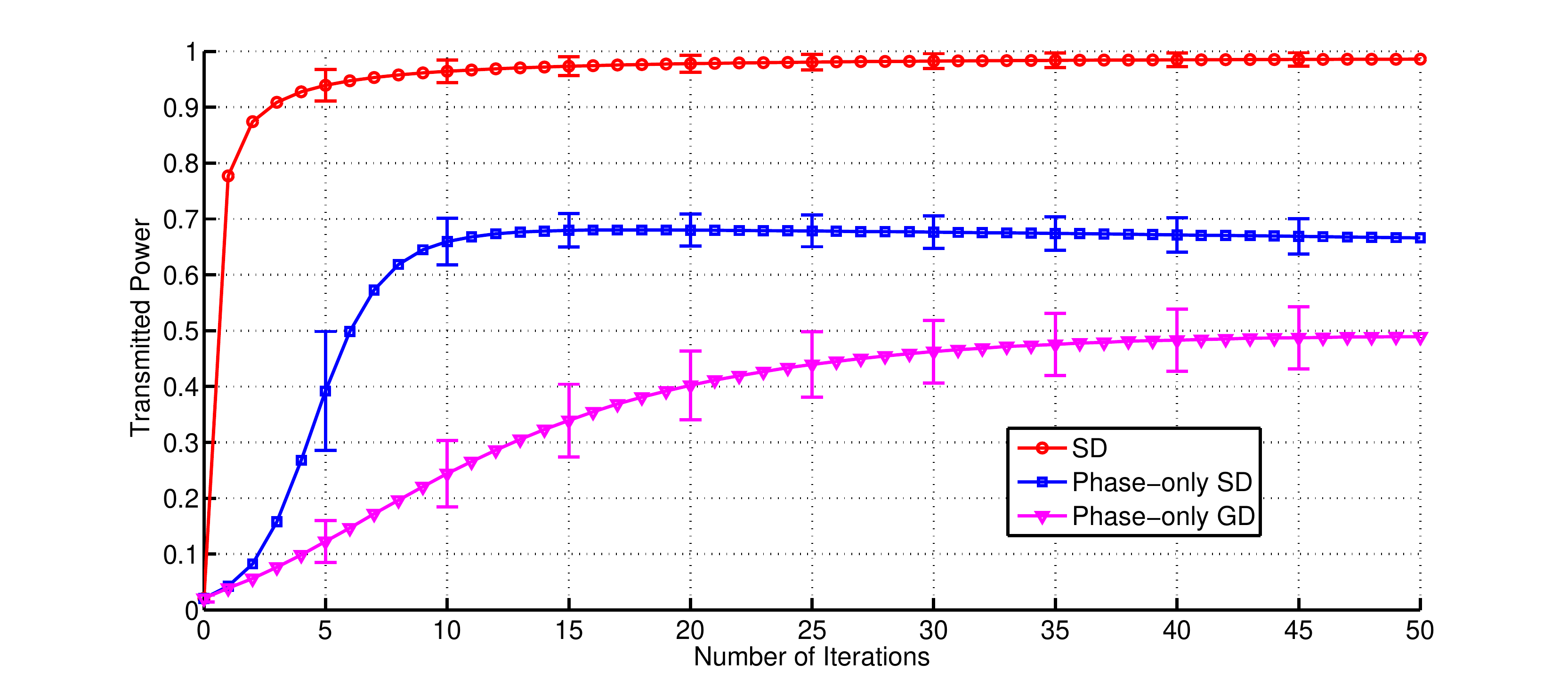}
\caption{The average transmitted power versus the number of iterations is shown for steepest descent algorithm, the phase-only steepest descent algorithm and the phase-only gradient descent algorithm for setup described in Fig. \ref{fig:SVDSDP}. Here $L = 3.4 \times 10^5 \lambda$. For each of these algorithms the optimal step size $\mu$ was chosen by a line search.}\vspace{-0.05cm}
\label{fig:opt convergence}
\end{figure}

 \color{black}
Fig.  \ref{fig:opt convergence} compares the rate of convergence of the phase-only modulated steepest descent and gradient descent based algorithms and the rate of convergence of the amplitude and phase modulated steepest descent based algorithm from \cite[Algorithm 1]{cjin2013}. Here we are in a setting with $D=197\lambda, L=3.4\times10^{5}\lambda, r=0.11\lambda, N_{c} = 430,000 \mbox{ dielectric cylinders with } n_{d} = 1.3, M = 395, \overline{l}=6.69\lambda$. The amplitude and phase modulated steepest descent algorithm produces a wavefront that converges to $95\%$ of the near optimum in about $5-10$ iterations as shown in Fig. \ref{fig:opt convergence}. The phase-only modulated steepest descent algorithm yields a highly transmitting wavefront within $5-10$ iterations. The phase-only modulated gradient descent algorithm also increases in transmission and converges in $15-20$ iterations. The fast convergence properties of the steepest descent based method make it suitable for use in an experimental setting where it might be infeasible to measure the $S_{11}$ matrix first.

\begin{figure}
\centering
\includegraphics[ trim = 0 5 0 5, clip = true,width=\textwidth]{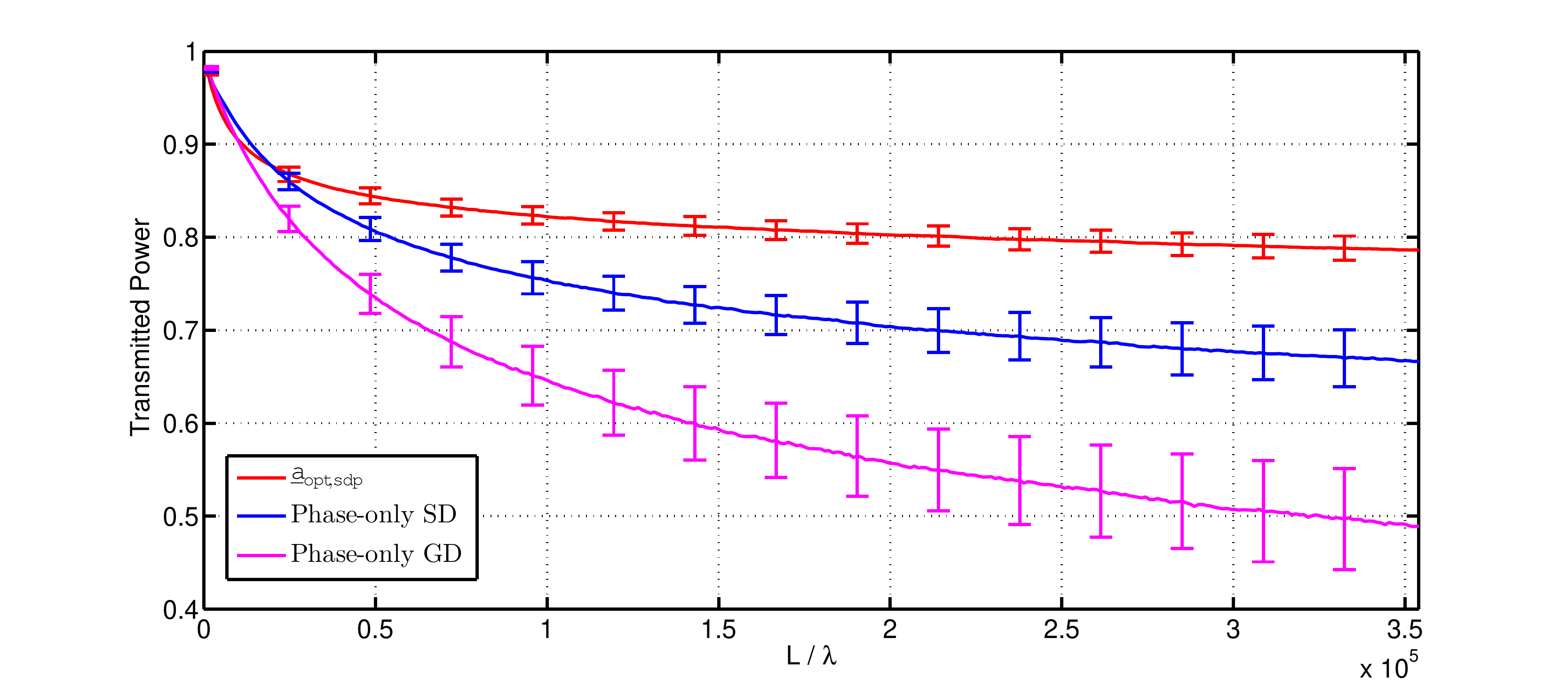}
\caption{The average transmitted power obtained after $50$ iterations of the phase-only steepest descent (SD) and gradient descent (GD) methods as a function of $L/\lambda$ for the setup described in Fig. \ref{fig:SVDSDP}. For comparison, we plot the transmitted power realized by $\aoptsdphat$. }\vspace{-0.05cm}
\label{fig:IMComp}
\end{figure}

Fig.  \ref{fig:IMComp} compares the maximum transmitted power achieved after $50$ iterations as a function of thickness $L/\lambda$  for the iterative, phase-only modulated steepest descent and gradient descent methods and the non-iterative SVD and SDP methods. The non-iterative methods increase transmission by $8.3\%$ relative to the steepest descent method. The gradient descent method performs poorly relative to the steepest descent method but still achieves increased transmission relative to the non-adaptive `equal-phase' wavefront.

\begin{figure}
\centering
\includegraphics[width=\textwidth]{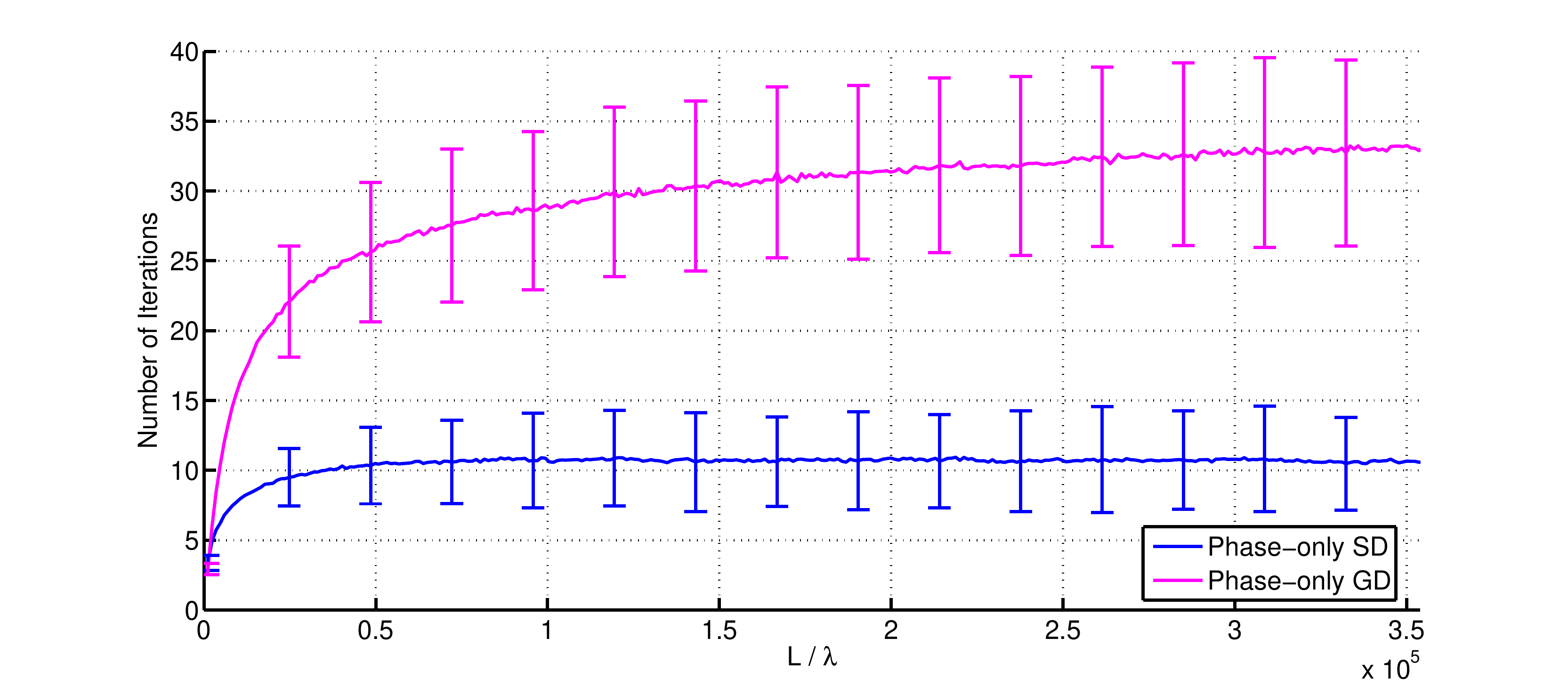}
\caption{Average number of iterations to get to $95\%$ of the respective maximum transmitted power for the phase-only steepest descent and gradient descent algorithms as a function of $L/\lambda$ for the setup described in Figure \ref{fig:SVDSDP}.}\vspace{-0.05cm}
\label{fig:NoIters}
\end{figure}

Fig. \ref{fig:NoIters} plots the average number of iterations required to reach $95\%$  of the respective optimas for the phase-only modulated steepest descent and gradient descent algorithms as a function of the thickness $L/\lambda$ of the scattering system. On average the steepest descent algorithm converges in about  in about $15-20$ iterations while the gradient descent algorithm converges in about $35-45$ iterations.

As the steepest descent algorithm converges faster and realizes $15-20\%$ greater transmitted power, but only loses $~10\%$ transmission relative to the non-iterative phase-only modulated SVD and SDP algorithms, it is the best option for use in an experimental setting.

\begin{figure}
\centering
\includegraphics[trim = 0 5 0 5, clip = true, width=5.55in]{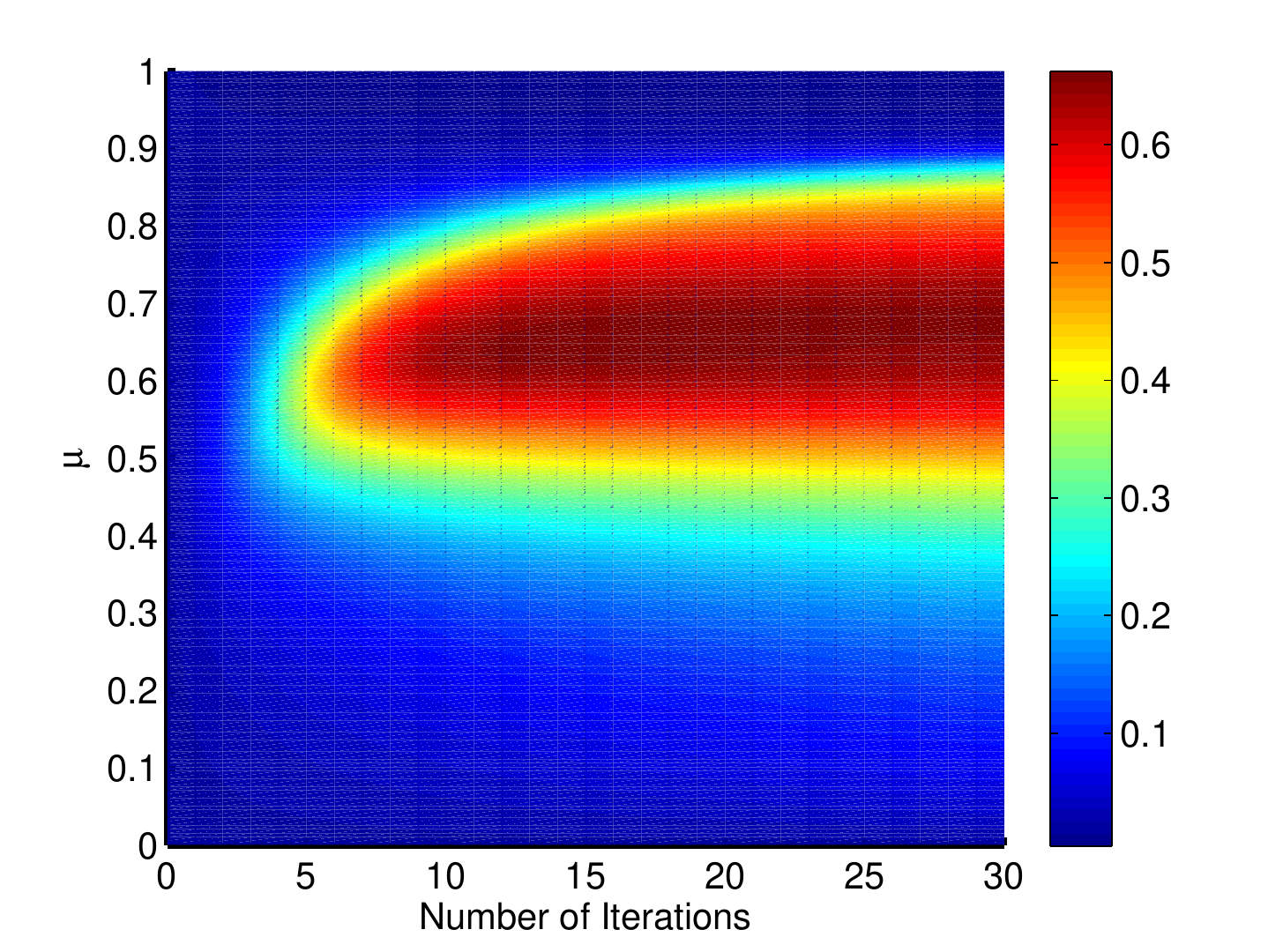}
\caption{Heatmap of the average transmitted power attained by the phase-only steepest descent algorithm as a function of the number of iterations and stepsize $\mu$ for the same setup as in Fig. \ref{fig:opt convergence}. }\vspace{-0.05cm}
\label{fig:MuSensitivitySDHeat}
\end{figure}

\begin{figure}
\centering
\includegraphics[trim = 0 5 0 25, clip = true, width=\textwidth]{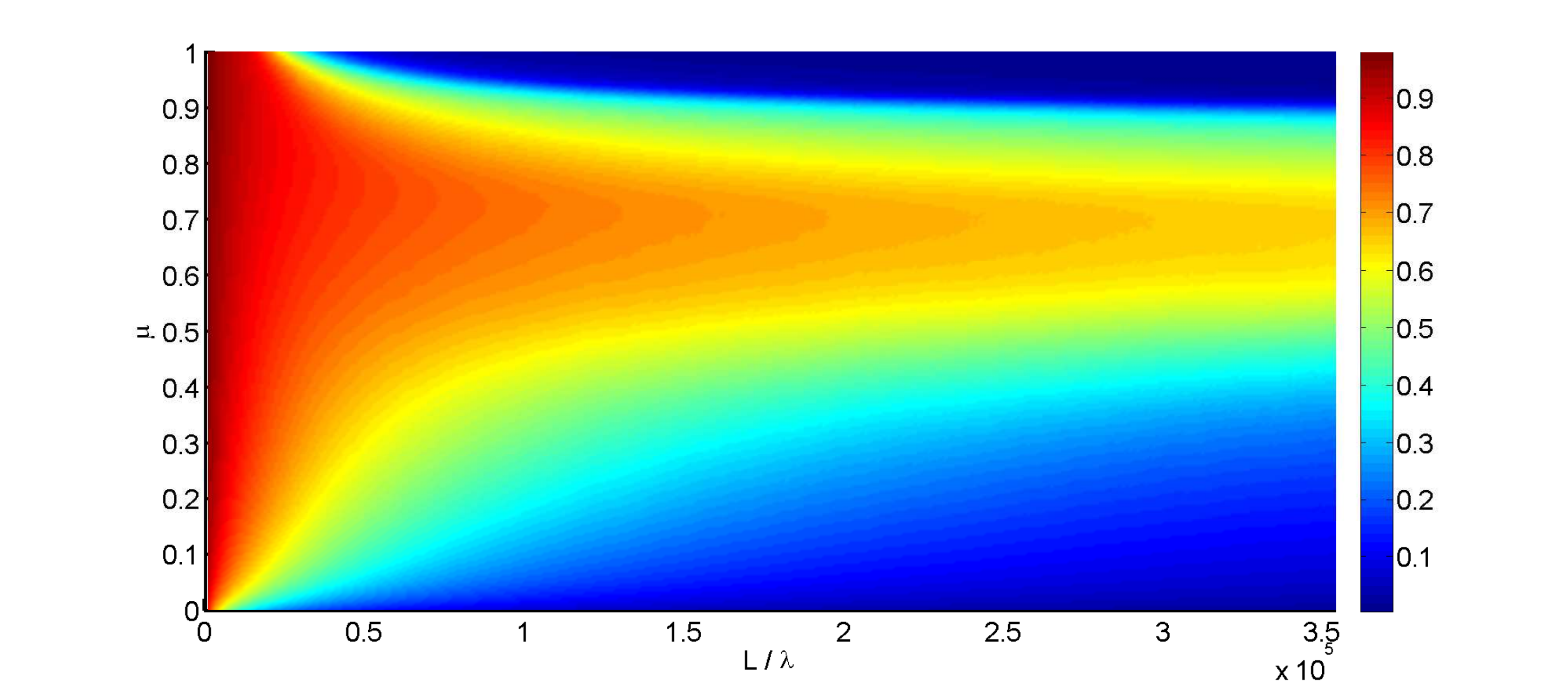}
\caption{Heatmap of the maximum transmitted power in 50 iterations of steepest descent on the plane of stepsize and the thickness $L/\lambda$ for the setup in Figure \ref{fig:SVDSDP}.}\vspace{-0.05cm}
\label{fig:MuRangeSD}
\end{figure}

Since determining the optimal size $\mu$ via a line search increases the number of measurements, we now investigate the sensitivity of the phase-only  steepest descent algorithm to the choice of stepsize. Fig.  \ref{fig:MuSensitivitySDHeat} plots the average transmitted power as a function of the number of iterations and the stepsize $\mu$ for the steepest descent algorithm.  This plot reveals that there is a broad range of $\mu$ for which the algorithm converges rapidly. Fig.  \ref{fig:MuRangeSD} shows the transmitted power achieved after $50$ iterations of the phase-only modulated steepest descent algorithm as a function of the stepsize $\mu$ and the thickness $L/\lambda$ of the scattering system showing that there is a wide range of allowed values for $\mu$ for which the steepest descent algorithm performs well. We have experimentally found that setting $\mu \approx 0.65$ yields fast convergence about $15-20$ iterations under a broad range of conditions.

\begin{figure}
\centering
\includegraphics[width=5.55in]{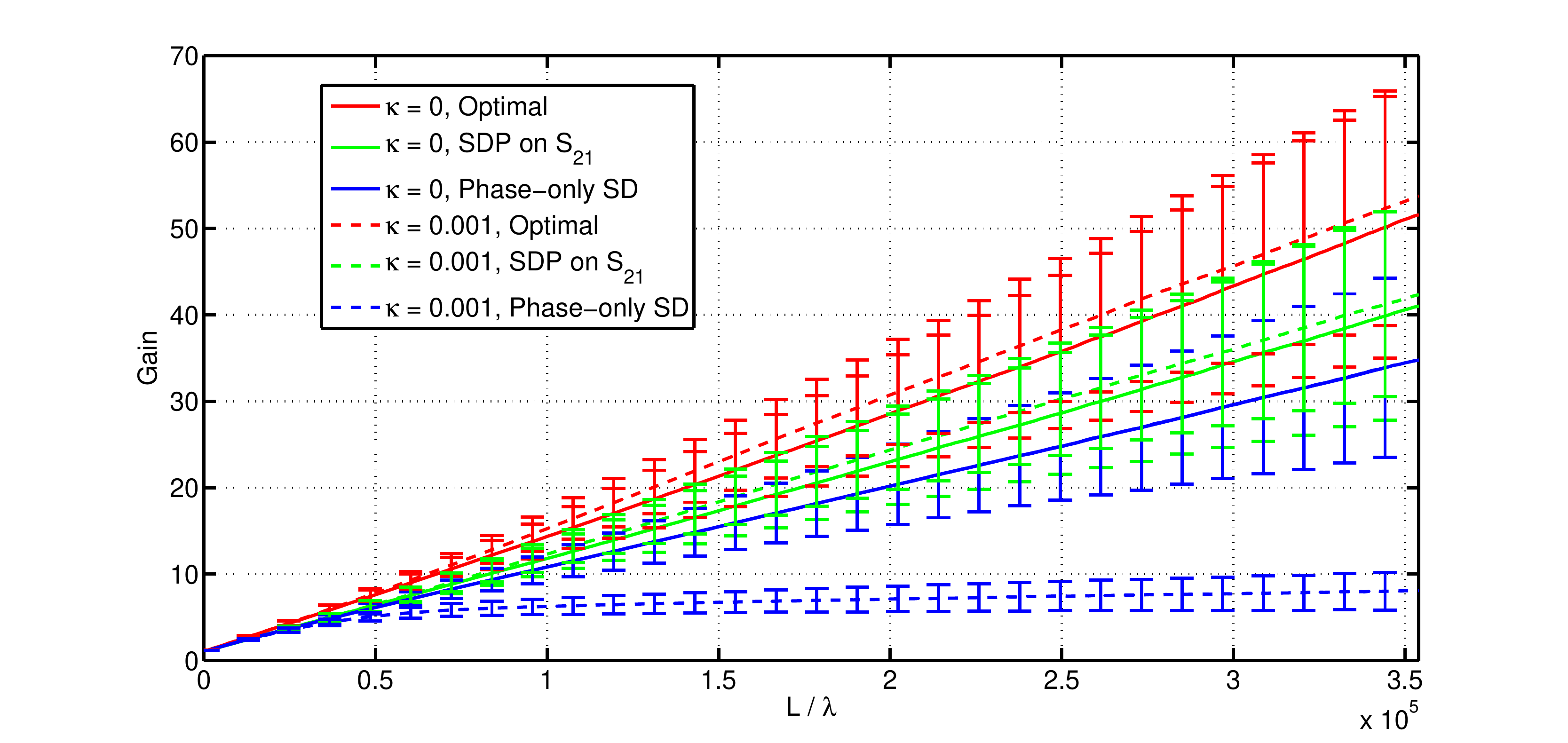}
\caption{Gain in transmitted power relative to $\aequal$ versus thickness $L/\lambda$ for a system setup as described in Fig. \ref{fig:SVDSDP} except with $n_{d} = 1.3 - j\kappa$, where $\kappa$ is the extinction coefficient.}\vspace{-0.05cm}
\label{fig:Gain Lossy}
\end{figure}

\color{black}
Finally, we consider the setting where the scatterers are absorptive with a refractive index given by $n_{d} = 1.3-j\kappa$. Here, backscatter minimization as a general principle for increasing transmission is clearly sub-optimal since an input with significant absorption can also minimize backscatter. In Fig.  \ref{fig:Gain Lossy}, we compare the gain, relative to $\tequal$, of the transmitted power achieved by the iterative phase-only steepest descent algorithm and non-iterative algorithms that assume knowledge of the $S_{21}$ matrix. Specifically, we compare the transmissions achieved by the wavefront produced by the backscatter analysis based steepest descent algorithm, the optimal transmission maximizing wavefront $\underline{v}_{1}$ which requires amplitude and phase modulation and the wavefront $\aoptsdphat$ obtained as in Eq. (\ref{eq:aoptsdp}), except with $A_{\sf sdp}$ defined as the solution of the optimization  problem
\begin{equation}\label{eq:SDP oracle}
\begin{aligned}
A_{\sf sdp} =& \argmax_{A \in \mathbb{C}^{M \times M}} \Trace\left( S_{21}^{H} \cdot S_{21} \cdot A \right)  \\
& \text{subject to }   A= A^H, A \succeq 0, \textrm{ and }  A_{ii} = 1/M \textrm{ for } i = 1, \ldots M.\\
\end{aligned}
\end{equation}
Fig. \ref{fig:Gain Lossy} shows that the iterative method realizes a significant increase in transmission even when the scatterers are weakly absorptive. The iterative algorithm converges rapidly, in about as many iterations as in the lossless setting for the same range of stepsizes

\section{Conclusions}\label{sec:conclusions}
We have shown theoretically and using numerically rigorous simulation that non-iterative, phase-only modulated techniques for transmission maximization using backscatter analysis can expect to achieve about $25 \, \pi \% \approx  78.5\%$ transmission in highly backscattering random media in the DMPK regime where amplitude and phase modulated can yield $~ 100 \%$ transmission. We have developed two new, iterative and physically realizable algorithms for constructing highly transmitting phase-only modulated wavefronts using backscatter analysis. We showed using numerical  simulations that the steepest descent variant outperforms the gradient descent variant and that the wavefront produced by the steepest descent algorithm achieves about $71\%$ transmission while converging within $15-20$ measurements.  The development of iterative phase-only modulated algorithms that bridge the $10\%$ transmission gap between the steepest descent algorithm presented here and the non-iterative SVD and SDP algorithms remains an important open problem. \color{black} We would also like to theoretically analyze the convergence properties of the iterative methods presented so we might better understand why the physically realizable variant of gradient descent method performs poorly compared to the physically realizable variant of the steepest descent method.
\color{black}

The proposed algorithms are quite general and may be applied to scattering problems beyond the 2D setup described in the simulations. A detailed study, guided by the insights in \cite{choi2011transmission},  of the impact of periodic boundary conditions on the results obtained is also underway.

\section*{Acknowledgements}

This work was partially supported by an ONR Young Investigator Award N000141110660, an NSF grant CCF-1116115, an AFOSR Young Investigator Award FA9550-12-1-0266 and an AFOSR DURIP Award FA9550-12-1-0016. We thank Jeff Fessler for suggesting the gradient descent method so that the advantage of the steepest descent based method could be properly showcased.

\setcounter{equation}{0}
\renewcommand{\theequation}{A{\arabic{equation}}}

\appendix

\section{Solving Eq. (\ref{eq:SDP problem}) in MATLAB}\label{sec:solvers}

Specifically, the solution to Eq. (\ref{eq:SDP problem}) can be computed in MATLAB using the CVX package \cite{cvx, gb08} by invoking the following sequence of commands:
\begin{verbatim}
cvx_begin sdp
    variable A(M,M) hermitian
    minimize trace(S11'*S11*A)
    subject to
    A >= 0;
    diag(A) == ones(M,1)/M;
cvx_end
Asdp = A; % return optimum in variable Asdp
\end{verbatim}

For settings where $M > 100$, we recommend using the SDPT3  solver \cite{SDPT3}. The solution to Eq. (\ref{eq:SDP problem}) can be computed in MATLAB using the SDPT3 package by invoking the following sequence of commands:

\begin{verbatim}
cost_function = S11'*S11;
e = ones(M,1); b = e/M;
num_params = M*(M-1)/2;
C{1} = cost_function;
A = cell(1,M); for j = 1:M, A{j} = sparse(j,j,1,M,M); end
blk{1,1} = 's';  blk{1,2} = M; Avec = svec(blk(1,:),A,1);
[obj,X,y,Z] = sqlp(blk,Avec,C,b);
Asdp = cell2mat(X);  % return optimum in variable Asdp
\end{verbatim}

\section{Derivation of Eq. (\ref{GD:gradient})}
\label{app:GDGradient}

Here, we derive Eq. (\ref{GD:gradient}). For notational brevity, we replace $S_{11}$ with $B$, and denote $B$'s $m$th row and $n$th column element as $B_{mn}$. We will show that

\begin{equation} \frac{\partial \| B \cdot \pmf(\pv) \|_{2}^{2} }{\partial \pv} = 2 \,\Imag\left[ \mbox{diag}\{ \pmf(-\pv) \} \cdot B^{H} \cdot B \cdot \pmf(\pv) \right]. \label{eq:required proof}
\end{equation}
To this end, note that the cost function can be expanded as
\begin{align}
    \nonumber \| B \cdot \pmf(\pv) \|_{2}^{2} &= \displaystyle \sum_{n=1}^{M} \left| B_{nm}e^{j\theta_m} \right|^{2}\\
    \nonumber &= \sum_{n=1}^{M} \sum_{m=1}^{M} \left| B_{nm} \right|^{2} + 2 \sum_{n=1}^{M} \sum_{p>q} \Real \left( B_{np} B_{nq}^{*} e^{j(\theta_p - \theta_q)} \right)\\
    &= \sum_{n=1}^{M} \sum_{m=1}^{M} \left| B_{nm} \right|^{2} + 2 \sum_{n=1}^{M} \sum_{p>q} |B_{np}||B_{nq}|\cos(\theta_p-\theta_q+\phase{B_{np}}-\phase{B_{nq}}),
\end{align}
where $\Real(\cdot)$ denotes the operator that returns the real part of the argument.\\
Consequently, the derivative of the cost function with respect to the $k$th phase $\theta_{k}$ can be expressed as
\begin{align}
    \dfrac{\partial \| B \cdot \pmf(\pv) \|_{2}^{2}}{\partial \theta_k} &= -2 \displaystyle \sum_{n=1}^{M} \sum_{q\neq k} \Imag \left[ B_{nk}B_{nq}^{*} e^{j(\theta_k-\theta_q)} \right]\\
    &= -2 \,\Imag \left[ e^{j\theta_k} \sum_{n=1}^{M} B_{nk} \sum_{q \neq k} B_{nq}^{*} e^{-j\theta_q} \right]  \label{eq:last step},
\end{align}
where $\Imag(\cdot)$ denotes the operator that returns the imaginary part of the argument.\\
Let  $\underline{e}_{k}$ be the $k$-th elementary vector. We may rewrite Eq. (\ref{eq:last step}) as
\begin{align}
    \dfrac{\partial \| B \cdot \pmf(\pv) \|_{2}^{2}}{\partial \theta_k} &= -2 \Imag\left[ e^{j\theta_k} \begin{bmatrix} B_{1k} & \cdots & B_{Mk} \end{bmatrix} \cdot B^{*} \cdot \left\{ I - \underline{e}_{k} \cdot \underline{e}_{k}^{H} \right\} \cdot \pmf(\pv)^{*} \right],
\end{align}
or, equivalently, as
\begin{align}
    \dfrac{\partial \| B \cdot \pmf(\pv) \|_{2}^{2}}{\partial \theta_k} &= -2 \Imag\left[ e^{j\theta_k} \begin{bmatrix} B_{1k} & \cdots & B_{Mk} \end{bmatrix} \cdot B^{*} \cdot \pmf(\pv)^{*} \right] -2 \Imag\left[  \begin{bmatrix} B_{1k} & \cdots & B_{Mk} \end{bmatrix} \cdot B^{*} \cdot \underline{e}_{k} \right]\\
    &= -2 \Imag\left[ e^{j\theta_k} \begin{bmatrix} B_{1k} & \cdots & B_{Mk} \end{bmatrix} \cdot B^{*} \cdot \pmf(\pv)^{*} \right].
\end{align}
Stacking the elements into a vector yields the relation
\begin{align}
    \dfrac{\partial \| B \cdot \pmf(\pv) \|_{2}^{2}}{\partial\pv} &= -2 \, \Imag\left[ \mbox{diag}\{ \pmf(\pv) \} \cdot B^{T} \cdot B^{*} \cdot \pmf(\pv)^{*} \right],
\end{align}
or, equivalently, Eq. (\ref{eq:required proof}).

\bibliographystyle{plain}
\bibliographystyle{unsrt}
\end{document}